\documentclass[journal]{IEEEtran}
\usepackage[utf8]{inputenc} 
\usepackage{graphicx}
\usepackage{setspace}
\usepackage{url}
\usepackage{amsmath,amssymb}
\usepackage[thmmarks,amsmath]{ntheorem}
\usepackage{xcolor}
\usepackage{balance}
\usepackage{subfigure}
\usepackage{upgreek}
\usepackage{cite}
\usepackage{algorithm,float}
\usepackage{algorithmicx}
\usepackage{algpseudocode}
\newtheorem{lemma}{Lemma}
\newtheorem{remark}{Remark}
\newtheorem{definition}{Definition}
\newtheorem{proposition}{Proposition}
\newtheorem{theorem}{Theorem}
\newtheorem{corollary}{Corollary}
\newtheorem{assumption}{Assumption}
\newtheorem{example}{Example}

\begin{document}
\title{Convergence Analysis of \\Signed Nonlinear Networks}

\author{Hao Chen,~~Daniel Zelazo,~~Xiangke Wang,~~and Lincheng Shen
\thanks{The work of H. Chen was supported in part by China Scholarship Council, in part by Hunan Provincial Innovation Foundation for Postgraduate (grant No. CX2017B014), and in part by National Natural Science Foundation of China (grant No. 61603406 and No. 61702528).  The work of D. Zelazo was supported by the Israel Science Foundation (grant No. 1490/1).	\emph{(Corresponding author: Daniel Zelazo.)}}
\thanks{H. Chen, X. Wang and L. Shen are with the College of Intelligence Science and Technology, 
	National University of Defense Technology, Changsha, 410073, China (e-mail: chenhao09@nudt.edu.cn; xkwang@nudt.edu.cn; lcshen@nudt.edu.cn).} 
\thanks{D. Zelazo is with the Faculty of Aerospace Engineering, 
	Technion-Israel Institute of Technology, Haifa, 32000, Israel. (email: dzelazo@technion.ac.il).} 
}
\maketitle

\begin{abstract}
	This work analyzes the convergence properties of signed networks with nonlinear edge functions.
	We consider diffusively coupled networks comprised of maximal equilibrium-independent passive (MEIP) dynamics on the nodes, and a general class of nonlinear coupling functions on the edges.  The first contribution of this work is to generalize the classical notion of signed networks for graphs with scalar weights to graphs with nonlinear edge functions using notions from passivity theory.  
	We show that the output of the network can finally form one or several steady-state clusters if all edges are positive, and in particular, all nodes can reach an output agreement if there is a connected subnetwork spanning all nodes and strictly positive edges.
%
	%
	When there are non-positive edges added to the network, 
	we show that the tension of the network still converges to the equilibria of the edge functions if the relative outputs of the nodes connected by non-positive edges converge to their equilibria.
	%
	%
	Furthermore, we establish the equivalent circuit models for signed nonlinear networks, and define the concept of equivalent edge functions which is a generalization of the notion of {effective} resistance. 
	We finally characterize the relationship between the convergence property and the equivalent edge function, when a non-positive edge is added to a strictly positive network comprised of nonlinear integrators.
	%
	%
	We show that the convergence of the network is always guaranteed, if the sum of the equivalent edge function of the previous network and the new edge function is passive.
\end{abstract}
\begin{IEEEkeywords}
	Nonlinear circuits, nonlinear networks, passivity, signed networks.
\end{IEEEkeywords}
\section{Introduction}
%
%
%
%
%

A common theme in many works on the multi-agent system is that the interaction between agents is \emph{cooperative}.  That is, the weights in the interaction protocol are positive.  There has been recent interest in protocols where the interactions may be either cooperative or antagonistic, with antagonistic interactions modeled by negative weights in the protocol.  Networks modeled by graphs with both positive and negative edge weights are termed \emph{signed networks}~\cite{Harary1953,Altafini2013}.
Signed networks have been studied in social network analysis~\cite{yang2007community,Xia2016} and multi-robot coordination~\cite{Qin2017On}, to name a few. 
As shown in~\cite{Zelazo2017}, in a signed network, agents may reach agreement on a common value (i.e., \emph{consensus}), form \emph{clusters}, or even diverge. 

%
The definiteness of the signed Laplacian matrix is a powerful tool for the convergence analysis of such signed networks~\cite{Zelazo2017}.
%
%
%
In particular, the work of~\cite{bronski2014spectral} provided the bounds on the number of positive and negative eigenvalues of the signed Laplacian matrix.
The work of~\cite{PanLulu2016} related the number of negative eigenvalues of the signed Laplacian matrix to the number of negative edges in the network.
In~\cite{Zelazo2014}, a necessary and sufficient condition was given for the signed Laplacian matrix becoming indefinite when one edge weight is negative.  It was shown that the absolute value of the negative edge weight must be larger than the inverse of the effective resistance between the two nodes connected by the edge.
The same result was rederived in~\cite{ChenYongxin2016}, where two alternative proofs were provided based on geometrical and passivity-based approaches.
The definiteness of the signed Laplacian matrix for the signed directed networks has also been studied in~\cite{Ahmadizadeh2017,Mukherjee2018}.
The analysis 
in~\cite{Zelazo2017,Mukherjee2018} are both developed based on the \emph{edge agreement} framework, which is established in~\cite{zelazo2011edge} to investigate the convergence property of the network by analyzing the relative outputs of the nodes connected by each edge.
%
%
%

Currently, most of the literature above on the convergence analysis of signed networks are restricted to linear systems, that is, the edge weights are all scalars and the input-output (I/O) relationship on each edge is a linear function of the entire state of the network.
In many applications, nonlinear protocols are designed to achieve the desired behavior of the network.
For example, the celebrated Kuramoto model is often used to analyze the synchronization of coupled phase oscillators~\cite{acebron2005kuramoto}, and it has been shown that nonlinear functions including the Kuramoto model are well suited into the edge agreement framework formulated in~\cite{zelazo2011edge} by passivity analysis.
%
In~\cite{WangLong2010}, another typical nonlinear consensus protocol was proposed to achieve finite-time consensus.
However, discussions on the consensus of signed nonlinear networks have not received much attention yet. 
There are some discussions of the nonlinear consensus protocols of signed networks in~\cite{Altafini2013,Mathias2017}, but these protocols do not operate on the relative outputs of the node dynamics, which is the subject of this work. %
%
%

For the analysis of nonlinear networked systems, passivity theory plays an important role~\cite{Arcak2007,Mathias2014,Sharf2017,Jain2018}.
One variation of passivity, \emph{equilibrium independent passivity} (EIP) was proposed in~\cite{Hines2011}, which requires a system to be passive independent of its equilibrium, and the equilibrium I/O map is a single-valued \emph{function}.
An extension of EIP is \emph{maximal equilibrium independent passivity} proposed in~\cite{Mathias2014}, which relaxes the equilibrium I/O map to be \emph{relations}, instead of \emph{functions}.
%
%
%
Motivated by the literature, in this paper, we will investigate the convergence properties of a signed nonlinear network of MEIP nodes.
By using notions from passivity theory, we generalize the consensus results of single integrators in a signed linear network discussed in~\cite{Zelazo2017} to the case of MEIP nodes in a signed nonlinear network.
The main contributions of this paper can be summarized as follows:
\begin{itemize}
	\item[i)] We generalize the definition of signed linear networks to graphs with nonlinear functions on the edges. 
	\item[ii)] We show that for a positive network of MEIP nodes, convergence is always guaranteed and the outputs form one or several clusters. In particular, all nodes can reach an output agreement if there is a connected spanning subnetwork of all nodes and strictly positive edges. 
	\item[iii)] For networks comprised of nonlinear integrator agents, we show a connection to notions from electrical circuit theory and the equivalent circuit model to derive convergence results for networks with non-positive edges.  We also propose an algorithm for constructing equivalent edge functions.
	%
\end{itemize}
\noindent The results above are also supported throughout the manuscript with illustrative numerical examples.

The rest of the paper is organized as follows. 
%
%
In Section~\ref{sec:signedMEIP}, we establish our network model, and generalize the classical definition of the signed networks to the nonlinear case based on passivity.
The convergence analysis of positive networks and signed networks are provided in Section~\ref{sec:convergence of signed}.
In Section~\ref{sec:single}, we establish a connection between the circuit theory and the signed networks.
The simulation result of a signed network of single integrators are offered in Section~\ref{sec:simulation}, and the concluding remarks are given in Section~\ref{sec:conclusion}.
%
\paragraph*{Preliminaries}
We use an undirected graph $\mathcal{G}=(\mathcal{V},\mathcal{E})$ to model a network of agents, where $\mathcal{V}$ and $\mathcal{E}$ denote the set of nodes and edges, respectively. 
%
%
If there is an edge connecting nodes $i$ and $j$, we say that node $i$ is a neighbor of node $j$, denoted as $i\in\mathtt{N}_j$.
In an undirected network, if $i\in\mathtt{N}_j$, then $j\in\mathtt{N}_i$. 
By assigning an arbitrary orientation to each edge, we can define the incidence matrix $E\in\mathbb{R}^{|\mathcal{V}|\times|\mathcal{E}|}$ as follows: $[E]_{ik}=1$, if edge $k$ is coming from node $i$; $[E]_{ik}=-1$, if edge $k$ is ending at node $i$; and $[E]_{ik}=0$ otherwise. 
%
For connected graphs it follows that the null space of $E^T$, denoted as $\mathcal{N}(E^T)=\beta\mathbf{1}$, where $\beta\in\mathbb{R}$~\cite{Godsil}. 
A \emph{directed path} from node $i$ to node $j$ in $\mathcal{G}$ is a subgraph of $\mathcal{G}$, where a sequence of edges connect a sequence of nodes, and the edges are oriented in the same direction from node $i$ to node $j$~\cite{Bondy1976}.
	Denote $\mathbf{P}_{i,j}$ as the set of all the paths from node $i$ to node $j$.
	When we say an edge $k\in\mathcal{G}$ is in a path $P_{i,j}\in\mathbf{P}_{i,j}$, we do not require the original orientation of $k$ is consistent with the direction of the path $P_{i,j}$.

We follow the convention by using italic letters for dynamic variables, e.g., $y(t)$, and using normal font letters to denote constant signals, e.g., $\mathrm{y}$.

\section{Signed Nonlinear Networks~\label{sec:signedMEIP}}
In this section, we formulate our network model, and generalize the concept of signed networks with nonlinear edge functions based on notions from passivity theory.
\subsection{The Network Model\label{sec:networkmodel}}
\begin{figure} [!b]
	\centering
	\includegraphics[height=1.4in]{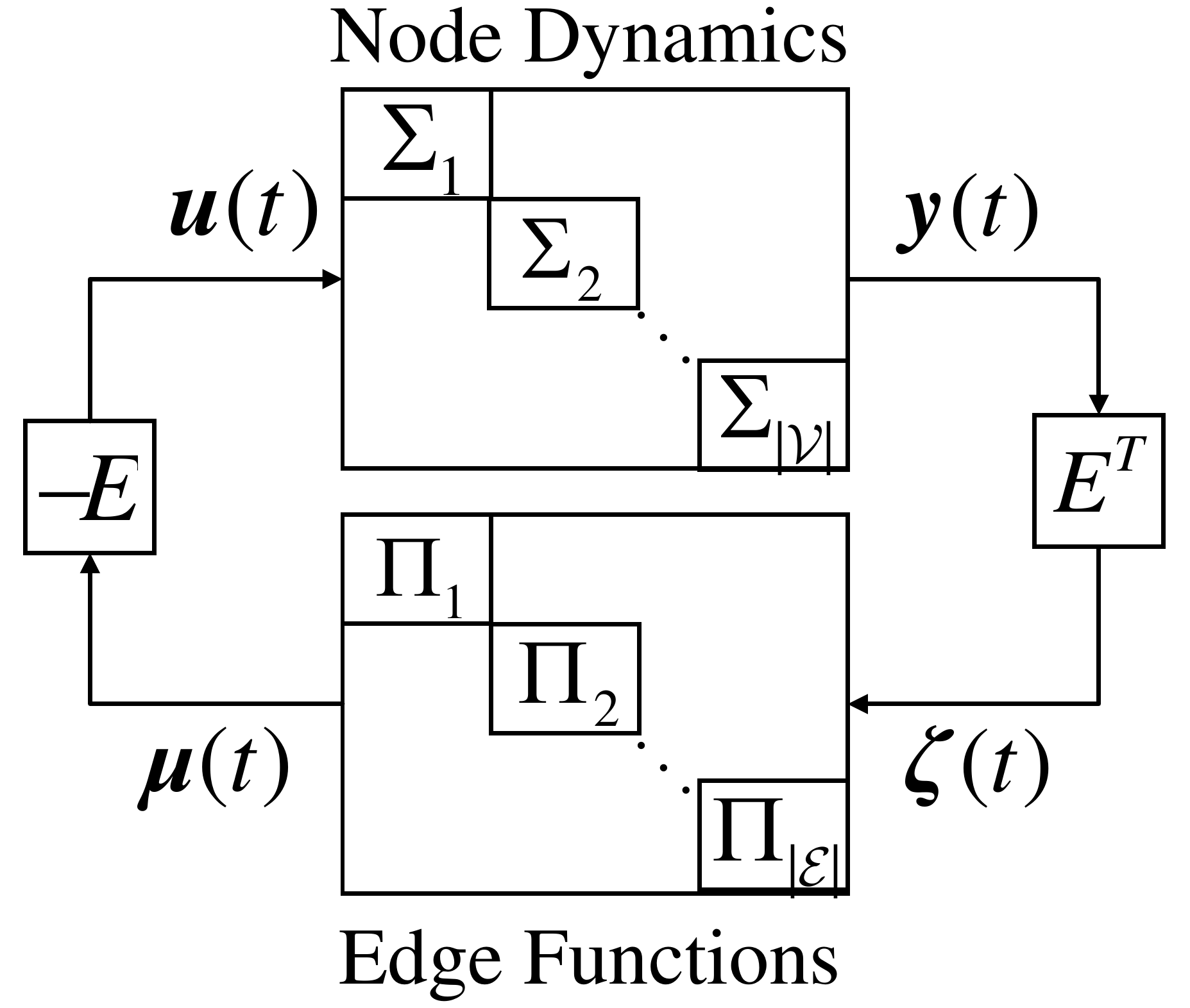}
	\caption{Network model of a collection of agents coupled with edge functions.}
	\label{fig_block}
\end{figure}

We present here the general model for our system and review an important extension of passivity theory for the analysis of these systems.  Consider a diffusively-coupled network of agents interacting over the graph $\mathcal{G}=(\mathcal{V},\mathcal{E})$, with each agent modeled as a single-input single-output (SISO) system represented by~\eqref{eq_plant},
\begin{equation}
\Sigma_i:~~\dot{x}_i(t)=f_i(x_i(t),u_i(t)),~~
y_i(t)=h_i(x_i(t),u_i(t)).
\label{eq_plant}
\end{equation}

We adopt the notation $\boldsymbol{u}(t)=[u_1(t),\ldots,u_{|\mathcal{V}|}(t)]^T$ and $\boldsymbol{y}(t)=[y_1(t),\ldots,y_{|\mathcal{V}|}(t)]^T$ for the stacked input and output vectors, respectively. Here, $x_i\in \mathcal{X}_i \subseteq \mathbb{R}^{p_i}$ is the system state, $u_i \in \mathcal{U}_i \subseteq \mathbb{R}$ the control input, and $y_i \in \mathcal{Y}_i \subseteq \mathbb{R}$ the system output.
%

%

Coordination among the agents is achieved by each agent interacting with its neighbors.
From the feedback control point of view, this can be modeled as a network of agents coordinated through the interactions on the edges. 
Each edge function utilizes the relative outputs between two adjacent nodes to generate the control signals, and the control signals are then added to both nodes to influence the nodes' internal states and outputs.
The block diagram of this network model is shown in Fig.~\ref{fig_block}, and the relative outputs of the nodes are defined with the incidence matrix as 
\begin{equation}
\label{eq_zetay}
\boldsymbol{\zeta}(t)=E^T\boldsymbol{y}(t),
\end{equation}
where $\boldsymbol{\zeta}(t)=[\zeta_1(t),\ldots,\zeta_{|\mathcal{E}|}(t)]^T$.
In many scenarios, we aim to let all nodes' outputs converge to the same value, i.e., $\boldsymbol{y}(t)\to\beta\mathbf{1},~\beta\in\mathbb{R}$ as $t \to \infty$, such that $\boldsymbol{\zeta}(t)\to\mathbf{0}$.
	This is the so-called \emph{consensus} (or agreement) problem.

On each edge $k\in\mathcal{E}$, there is a function  taking the relative output $\zeta_k(t)$ as input, and with the following form
\begin{equation}
\Pi_k:
\mu_k(t)=\psi_k(\zeta_k(t)).
\label{eq_edge}
\end{equation}
Similarly, we use the stacked vector form  $\boldsymbol{\mu}(t)=[\mu_1(t),\ldots,\mu_{|\mathcal{E}|}(t)]^T$ for the outputs of the edge functions.
The stacked version of~\eqref{eq_edge} can be represented as $\boldsymbol{\mu}(t)=\Psi(\boldsymbol{\zeta}(t))=[\psi_1(\zeta_1(t)),\ldots,\psi_{|\mathcal{E}|}(\zeta_{|\mathcal{E}|}(t))]^T$.
Denote $I_k$ as the set of equilibria of edge $k$. That is, $\psi_k(\zeta_k(t))=0$ if and only if $\zeta_k(t)\in I_k$.
%
We use $\boldsymbol{I}=I_1 \times I_2 \times \cdots \times I_{|\mathcal{E}|}$ to denote the equilibria of all the edge functions.
For each edge function $\psi_k(\cdot)$, we require $\psi_k(0)=0$, such that  $\boldsymbol{I}$ contains the origin, and $\boldsymbol{\mu}(t)\to\mathbf{0}$ when $\boldsymbol{\zeta}(t)\to\mathbf{0}$.
The edge output $\mu_k$ is added to the inputs of the two nodes that are connected by edge $k$.
Thus the relation between the stacked vector $\boldsymbol{u}(t)$ and $\boldsymbol{\mu}(t)$ can be represented as
\begin{equation}
\label{eq_umu}
\boldsymbol{u}(t)=-E\boldsymbol{\mu}(t).
\end{equation} 

%
Following equations~\eqref{eq_zetay} and~\eqref{eq_umu}, we have $\boldsymbol{u}(t)^T\boldsymbol{y}(t)=-\boldsymbol{\mu}(t)^TE^T\boldsymbol{y}(t)=-\boldsymbol{\mu}(t)^T\boldsymbol{\zeta}(t)$.

\begin{remark}
	\label{rmk_circuit}
	In~\cite{Mathias2014}, the nodes' outputs $\boldsymbol{y}(t)$ are named the \emph{potential}, the relative outputs of the nodes $\boldsymbol{\zeta}(t)$ the \emph{tension}, $\boldsymbol{\mu}(t)$ is called the \emph{flow}, and the node inputs $\boldsymbol{u}(t)$ as the divergence.
	Equation~\eqref{eq_zetay} is called Kirchhoff's voltage law (KVL), and equation~\eqref{eq_umu} is called Kirchhoff's current law (KCL).
	These concepts are borrowed from electrical circuit theory, which we discuss in Section~\ref{sec:single}. %
\end{remark}

Equations~\eqref{eq_plant}-\eqref{eq_umu} formulate a general framework of diffusively coupled systems, denoted by the triple $(\mathcal{G},\Sigma, \Pi)$.
According to \eqref{eq_plant}-\eqref{eq_umu}, $\boldsymbol{u}(t)=-E\Psi(E^T\boldsymbol{y}(t))$,
	$\Psi(\cdot)$ can be taken in accordance with the edge orientation, such that $\boldsymbol{u}$ is independent of the edge orientation.
\begin{remark}
When the node dynamics \eqref{eq_plant} are single integrators and the edge functions are linear (scalar) weights, this network model captures the celebrated consensus, or Laplacian, dynamics over graphs \cite{Mesbahi2010}.
\end{remark}

For nonlinear systems, passivity theory has emerged as a powerful tool for the convergence analysis of diffusively coupled networks~\cite{Arcak2007,Mathias2014,Sharf2017}.
In this work, we employ the notion of \emph{maximal equilibrium-independent passivity} (MEIP) in~\cite{Mathias2014}, 
which is an extension to results on equilibrium-independent passivity (EIP), originally proposed in~\cite{Hines2011}.  The key concept in both MEIP and EIP is to require that a passivity inequality holds between any system trajectory and forced equilibrium points.  The system theoretic machinery needed to apply these passivity notions is the characterization of \emph{equilibrium input-output maps}.  
In this direction, we assume that there exists a nonempty set $\bar{\mathcal{U}}_i \subseteq \mathcal{U}_i$ such that for every constant ${\rm u}_i \in \bar{\mathcal{U}}_i$, there exists a constant ${\rm y}_i \in \mathcal{Y}_i$. Define $\sigma_i$ as the I/O map which contains the set of all the steady-state I/O pairs $({\rm u}_i, {\rm y}_i)$. For MEIP systems, the I/O maps $\sigma_i$ are set-valued maps (or curves in $\mathbb{R}^2$), i.e., $\sigma_i = \{({\rm u}_i, {\rm y}_i): {\rm u}_i \in \bar{\mathcal{U}}_i\}$, and we denote ${\rm y}_i\in \sigma_i({\rm u}_i)$ if $({\rm u}_i, {\rm y}_i) \in \sigma_i$. 
The relation $\sigma_i$ is said to be \emph{maximally monotone} if $({\rm u}'_i, {\rm y}'_i), ({\rm u}''_i, {\rm y}''_i) \in \sigma_i$ then either $({\rm u}'_i \leq {\rm u}''_i$ and ${\rm y}'_i \leq {\rm y}''_i)$, or $({\rm u}'_i \geq {\rm u}''_i$ and ${\rm y}'_i \geq {\rm y}''_i)$, and $\sigma_i$ is not contained in any larger monotone relation \cite{Rockafellar1998}.
%
%
%
With the notion of \emph{maximal monotone}, we now define \emph{maximal equilibrium-independent passivity} (MEIP).
\begin{definition}[\cite{Mathias2014}]
	\label{def_meip}
	System $\Sigma_i$ represented by~\eqref{eq_plant}
	is said to be \emph{maximal equilibrium-independent passive}, if there exists a maximal monotone relation $\sigma_{i}$ such that for all equilibrium I/O relations $(\mathrm{u}_i,\mathrm{y}_i)\in \sigma_{i}$, there exists a positive semi-definite storage function $S_i(x_i(t))$ satisfying
	\begin{equation}
	\label{eq_meip}
	\dot{S}_i\leq(u_i(t)-\mathrm{u}_i)(y_i(t)-\mathrm{y}_i).
	\end{equation}
\end{definition}

We use $\mathbf{u}$, $\mathbf{y}$, and $\boldsymbol{\sigma}(\cdot)$ for the stacked equilibrium inputs, outputs, and I/O relations of all the nodes, i.e., $\mathbf{y}=\boldsymbol{\sigma}(\mathbf{u})$ means $\mathrm{y}_i\in \sigma_{i}(\mathrm{u}_i)$ for any $i\in \mathcal{V}$.  For more discussion and examples of MEIP systems, the reader is referred to \cite{Mathias2014}.
%

In this paper, we assume all the nodes are MEIP systems.
With the above formulation, when $\boldsymbol{y}(t)$ is in an agreement state (i.e., $\boldsymbol{y}(t)=\beta \mathbf{1}$) , the inputs to the nodes are $\boldsymbol{u}(t)=\mathbf{0}$.
To guarantee the existence of feasible equilibrium solutions corresponding to the agreement state, there should exist $\tilde{\mathbf{y}}\in\boldsymbol{\sigma}(\mathbf{0})\cap\mathcal{N}(E^T)$, such that the agreement space is an invariant set for $\boldsymbol{y}(t)$.
We put it into an assumption as follows:
\begin{assumption}
	\label{asp_node}
	Each node represented by~\eqref{eq_plant} is MEIP, with
	the equilibrium I/O relations satisfying $\boldsymbol{\sigma}(\mathbf{0})\cap\mathcal{N}(E^T)\neq\emptyset$.
\end{assumption}

Observe that if $\lim\limits_{t\rightarrow\infty}\boldsymbol{\zeta}(t)=\tilde{\boldsymbol{\upzeta}}$ exists, and $\tilde{\boldsymbol{\upzeta}}\in\boldsymbol{I}$, then $\lim\limits_{t\rightarrow\infty}\boldsymbol{\mu}(t)=\mathbf{0}$ and $\lim\limits_{t\rightarrow\infty}\boldsymbol{u}(t)=\lim\limits_{t\rightarrow\infty}-E\boldsymbol{\mu}(t)=\mathbf{0}$. If in addition Assumption~\ref{asp_node} holds, then $\lim\limits_{t\rightarrow\infty}\boldsymbol{y}(t)=\tilde{\mathbf{y}}$ exists, meaning the nodes have steady outputs as $t\to\infty$.
	If $\tilde{\mathbf{y}}=\beta\mathbf{1}$, $\beta\in\mathbb{R}$, it means the outputs of the nodes are finally in agreement;
	otherwise, it means the outputs of the nodes form multiple clusters, which is the so-called \emph{clustering} phenomenon.
	%
	%
	%

\subsection{Signed Nonlinear Edges\label{sec:signed}}
The study of signed networks has its origins in graph theory~\cite{Harary1953}.
The standard notion of signed networks considers graphs with edges labeled as either positive (+) or negative (-).  In the study of dynamic systems over graphs, such as the model considered here, the notion of signed networks relates to the sign of a scalar edge weight in a linear interaction protocol, defined below.
\begin{definition}[\cite{PanLulu2016}]
	\label{def_classicalsigned}
	Consider the edge function of the form
	\begin{equation}
	\label{eq_linear}
	\Pi_k:\mu_k(t)=w_k\zeta_k(t),
	\end{equation}	
	where $w_k \in \mathbb{R}$.  Then edge $k$ is \emph{positive} if $w_k>0$, and is \emph{negative} if $w_k<0$.
\end{definition}

Practically, the positive edge represents the cooperative, trustful or attractive relationship between the nodes, while the negative edge corresponds to the antagonistic, distrustful or repulsive interactions.
However, Definition~\ref{def_classicalsigned} only deals with networks with scalar edge weights.
In many applications, the relationship between the nodes can be much more complicated.
In this paper, we generalize the concept of signed edges to the nonlinear edge functions based on notions from passivity theory.  
In this direction, 
we first review the standard definition of a passive system.

\begin{figure} [!t]	
	\centering
	\subfigure[]
	{\includegraphics[height=.83in]{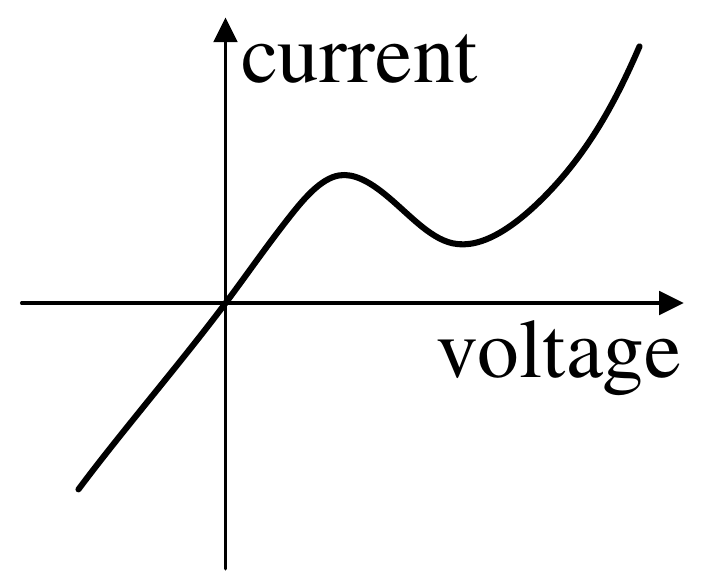}	
		\label{fig_tunnel}}
	\subfigure[]
	{\includegraphics[height=.83in]{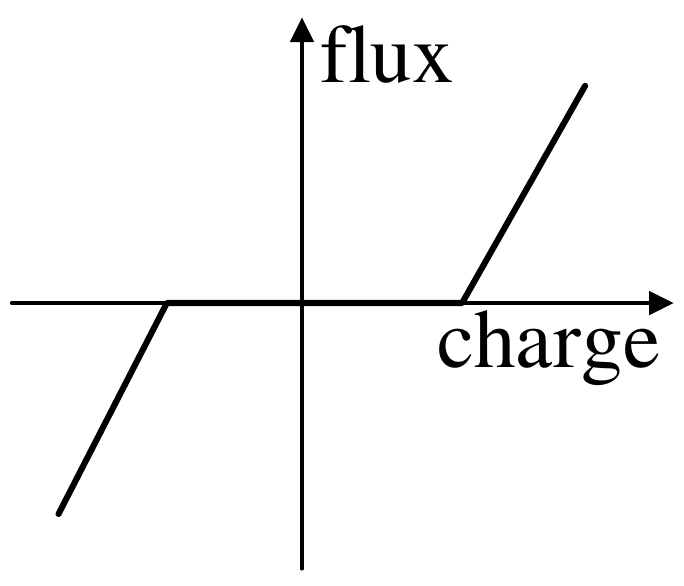}
		\label{fig_memristor}	}	
	\subfigure[]
	{\includegraphics[height=.83in]{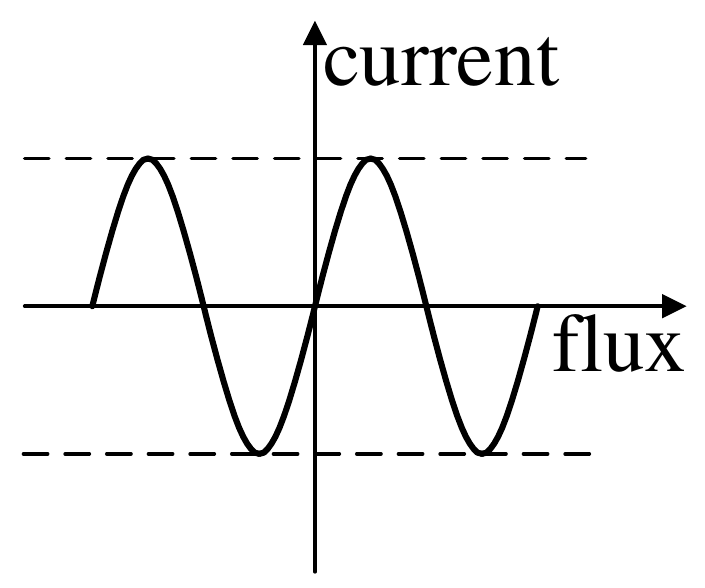}	
		\label{fig_josephson}}		
	\caption{Some real elements with nonlinear characteristics~\cite{Hasler}. (a) The current-volatage characteristic of a tunnel diode; (b) the flux-charge characteristic of a memristor; (c) the current-flux characteristic of a Josephson junction.
	}
	\label{fig_examplefunction}
\end{figure}

\begin{definition}[\cite{Hassan}]
	\label{def_passive}
	A system $\eta=\pi(t,\xi)$, where $\xi,\eta$ are the system input vector and system output vector, respectively, is
	\begin{itemize}
		\item[i)] \emph{passive} if $\xi^T\eta\geq0$,  $\forall(t,\xi)$;
		\item[ii)] \emph{input strictly passive} if $\xi^T\eta\geq\epsilon\xi^T\xi$, where $\epsilon>0$,  $\forall(t,\xi)$.
	\end{itemize}
\end{definition}
%
%

The linear edge function~\eqref{eq_linear} with $w_k>0$ is input strictly passive, and one can choose $\epsilon=w_k/2$ to arrive at that conclusion.
To account for general negative nonlinear edges, following Definition~\ref{def_passive}, we introduce the notion of \emph{active} and \emph{input strictly active} systems.
\begin{definition}
	\label{def_active}
	A system $\eta=\phi(t,\xi)$, where $\xi,\eta$ are system input vector and output vector, respectively, is
	\begin{itemize}
		\item[i)] \emph{active} if $\xi^T\eta\leq0$,  $\forall(t,\xi)$;
		\item[ii)] \emph{input strictly active} if $\xi^T\eta\leq-\epsilon\xi^T\xi$, where $\epsilon>0$,  $\forall(t,\xi)$.		
	\end{itemize}
\end{definition}

With Definition~\ref{def_passive} and~\ref{def_active} in place, we are now able to define a notion of signed nonlinear networks.
\begin{definition}
	\label{def_signed}
	Suppose the edge function~\eqref{eq_edge} is a map from $\mathbb{R}$ to $\mathbb{R}$, with $\psi_k(0)=0$. Then edge $k$ is termed  
	\begin{itemize}
		\item[i)] \emph{(strictly) positive} if~\eqref{eq_edge} is (input strictly) passive;
		\item[ii)] \emph{(strictly) negative} if~\eqref{eq_edge} is (input strictly) active.
	\end{itemize}
\end{definition}

%
The traditional definition of signed edges with scalar weights also fall under these categories.
An edge described by function~\eqref{eq_linear} is strictly positive if $w_k>0$, and is strictly negative if $w_k<0$.
The positive edges and negative edges in Definition \ref{def_signed} are broader in scope.  Such edge functions can have more than one equilibria besides the origin.
%
%
%

	Many real elements have such kind of nonlinear characteristics~\cite{Hasler}.
	%
	%
	Fig~\ref{fig_tunnel} shows the current-voltage characteristic of a \emph{tunnel diode}, whose curve has only one intersection with the voltage axis (at the origin), and the edges with such a function are stictly positive.
	Fig.~\ref{fig_memristor} shows the flux-charge characteristic of a \emph{memristor}, whose curve has multiple intersections with the charge axis, and the edges with such a function are positive, but not strictly positive.  This curve can also represent sensors with \emph{dead zones}.
	If there is a negative gain multiplying the functions in Fig~\ref{fig_tunnel} and Fig.~\ref{fig_memristor}, then the edges with the resulting functions are strictly negative and negative, respectively.
	Fig.~\ref{fig_josephson} shows the current-flux characteristic of a \emph{Josephson junction}, which is neither positive nor negative. 

%

In this paper, we use $\mathcal{E}_{\geq}$, $\mathcal{E}_>$, $\mathcal{E}_{\leq}$ and $\mathcal{E}_<$ to denote the set of edges that are positive, strictly positive, negative, and strictly negative, respectively.  Note that in general, $\mathcal{E}_> \subseteq \mathcal{E}_{\geq}$ and $\mathcal{E}_< \subseteq \mathcal{E}_{\leq}$.  We also use the notion of \emph{non-positive} (\emph{non-negative}) edges defined to be the complement in $\mathcal{E}$ of $\mathcal{E}_{\geq}$ ($\mathcal{E}_{\leq}$).
Similarly, the set of \emph{non-strictly positive} edges is the complement of the set of strictly positive edges.
Therefore, the set of non-positive edges is a subset of the set made up of non-strictly positive edges.
%
With the above definition for signed edges, we now generalize the definition of signed networks to the nonlinear cases. 
\begin{definition}[{\small Signed Nonlinear Networks}]
	A networked system $(\mathcal{G},\Sigma, \Pi)$ is a
	\begin{itemize}
		\item \emph{positive network} if all edges are positive (i.e., $\mathcal{E} = \mathcal{E}_{\geq}$);
		\item \emph{strictly positive network} if all edges are strictly positive (i.e., $\mathcal{E} = \mathcal{E}_>$);
		\item \emph{signed network} 
		if not all edges are positive (i.e., $\mathcal{E}\setminus\mathcal{E}_{\geq}\neq\emptyset$).
	\end{itemize}
\end{definition}
%
%


\section{Convergence of Signed Nonlinear Networks~\label{sec:convergence of signed}} 
We begin by analyzing positive networks, and then move forward to the cases of general signed nonlinear networks.

\subsection{Positive Networks}
We first show that convergence is always guaranteed in a positive network of MEIP nodes.
\begin{theorem}
	\label{thm_interval}
	Consider a positive network system $(\mathcal{G},\Sigma,\Pi)$ with connected graph $\mathcal{G}=(\mathcal{V},\mathcal{E}_{\geq})$ represented by~\eqref{eq_plant}-\eqref{eq_umu} and suppose Assumption~\ref{asp_node} holds.  Then $\lim\limits_{t\rightarrow\infty}\boldsymbol{\zeta}(t)=\tilde{\boldsymbol{\upzeta}}$ exists, and
	$\tilde{\boldsymbol{\upzeta}}\in \boldsymbol{I}$.
\end{theorem}
\begin{IEEEproof}
	Consider the Lyapunov function $V(\boldsymbol{x}(t))=\sum_{i=1}^{|\mathcal{V}|}S_i(x_i(t))$, where $S_i(x_i(t))$ is the storage function of node $i$.
	According to~\eqref{eq_meip}, we have
	 $$
	 \dot{V}=\sum_{i=1}^{|\mathcal{V}|}\dot{S}_i\leq(\boldsymbol{u}(t)-\mathbf{0})^T(\boldsymbol{y}(t)-\tilde{\boldsymbol{\mathbf{y}}})
	 =-\boldsymbol{\zeta}(t)^T\boldsymbol{\mu}(t).
	 $$
	 Since the network is positive, according to Definition~\ref{def_passive} and~\ref{def_signed}, $-\boldsymbol{\zeta}(t)^T\boldsymbol{\mu}(t)\leq0$, with the equality holds if and only if $\boldsymbol{\mu}(t)=\mathbf{0}$ and $\boldsymbol{\zeta}(t)\in\boldsymbol{I}$.
	 
	 %
	%
	By using LaSalle's invariance principle~\cite{Hassan}, we can conclude that the system will converge to the largest invariant set satisfying $\boldsymbol{\zeta}(t)^T\boldsymbol{\mu}(t)=\mathbf{0}$, meaning $\lim\limits_{t\rightarrow\infty}\boldsymbol{\mu}(t)=\mathbf{0}$, and $\lim\limits_{t\rightarrow\infty}\boldsymbol{\zeta}(t)\in\boldsymbol{I}$.
	As a result, $\lim\limits_{t\rightarrow\infty}\boldsymbol{u}(t)=\mathbf{0}$, and $\lim\limits_{t\rightarrow\infty}\boldsymbol{y}(t)=\tilde{\mathbf{y}}$, where $\tilde{\mathbf{y}}\in\boldsymbol{\sigma}(\mathbf{0})$, therefore $\lim\limits_{t\rightarrow\infty}\boldsymbol{\zeta}(t)=\tilde{\boldsymbol{\upzeta}}=E^T\tilde{\mathbf{y}}$ exists.
	%
\end{IEEEproof}

Theorem~\ref{thm_interval} means that for a positive network of MEIP nodes, the steady states of the outputs can form one or several \emph{clusters}.
If there is only one cluster, then it is exactly the consensus case.
%
%
In fact, if the network is strictly positive, then $\boldsymbol{I}$ is the origin.
As $\boldsymbol{\zeta}(t)\to\mathbf{0}$, 
$\boldsymbol{y}(t)$ will converge to the agreement space.
This is formulated as the following corollary.
\begin{corollary}
	\label{thm_strictly}
	Consider a strictly positive network system $(\mathcal{G},\Sigma,\Pi)$ with connected graph  $\mathcal{G}=(\mathcal{V},\mathcal{E}_>)$ represented by~\eqref{eq_plant}-\eqref{eq_umu} and suppose Assumption~\ref{asp_node} holds. Then 
	 $\lim\limits_{t\rightarrow\infty}\boldsymbol{\zeta}(t)=\mathbf{0}$, and $\lim\limits_{t\rightarrow\infty}\boldsymbol{y}(t)=\beta\mathbf{1},~\beta\in\mathbb{R}$.
\end{corollary}

In fact, we do not need all edges to be strictly positive in order to reach agreement, as shown in the following corollary.
%
%
%
%
\begin{corollary}
	\label{coro_positive+strictly}
	Consider a positive network system $(\mathcal{G},\Sigma,\Pi)$ with connected graph $\mathcal{G}=(\mathcal{V},\mathcal{E}_{\geq})$ represented by~\eqref{eq_plant}-\eqref{eq_umu} and suppose Assumption~\ref{asp_node} holds.
	If there exists a connected subgraph $\mathcal{G}_>=(\mathcal{V},\mathcal{E}_>)$ spanning all nodes and strictly positive edges, then 
	$\lim\limits_{t\rightarrow\infty}\boldsymbol{\zeta}(t)=\mathbf{0}$, and $\lim\limits_{t\rightarrow\infty}\boldsymbol{y}(t)=\beta\mathbf{1},~\beta\in\mathbb{R}$.
\end{corollary}
\begin{IEEEproof}
	According to Definition~\ref{def_passive} and~\ref{def_signed}, $I_k=0$, $\forall~k\in\mathcal{E}_>$.
	With Theorem~\ref{thm_interval}, we get $\lim\limits_{t\rightarrow\infty}\zeta_k(t)=0$, $\forall~k\in\mathcal{E}_>$, meaning $\lim\limits_{t\rightarrow\infty}y_i(t)-y_j(t)=0$, where $i,j\in \mathcal{V}$ are the two nodes connected by edge $k$.
	Since $\mathcal{G}_>=(V,\mathcal{E}_>)$ is connected, therefore, $\lim\limits_{t\rightarrow\infty}y_i(t)-y_j(t)=0$, $\forall~i,j\in \mathcal{V}$.
	As a result, $\lim\limits_{t\rightarrow\infty}\boldsymbol{\zeta}(t)=\lim\limits_{t\rightarrow\infty}E^T\boldsymbol{y}(t)=\mathbf{0}$, and $\lim\limits_{t\rightarrow\infty}\boldsymbol{y}(t)=\tilde{\mathbf{y}}$, where $\tilde{\mathbf{y}}\in\boldsymbol{\sigma}(\mathbf{0})\cap\mathcal{N}(E^T)$, i.e., $\lim\limits_{t\rightarrow\infty}\boldsymbol{y}(t)=\beta\mathbf{1},~\beta\in\mathbb{R}$.
\end{IEEEproof}

	In the case where steady-state clusters are formed in positive networks, we are also able to provide bounds on the distance between these clusters. 
	Suppose the equilibria of edge $k$ is contained in a closed interval $[I_k^L,I_k^R]$, with $I_k^L\leq0$, and $I_k^R\geq0$, i.e., $I_k\subset[I_k^L,I_k^R]$.
	%
	We have the following corollary indicating the bounds of the distances between the steady outputs of any pair of nodes.
	\begin{corollary}
		\label{coro_distance}
		Consider a positive network system $(\mathcal{G},\Sigma,\Pi)$ with connected graph $\mathcal{G}=(\mathcal{V},\mathcal{E}_{\geq})$ represented by~\eqref{eq_plant}-\eqref{eq_umu} and suppose Assumption~\ref{asp_node} holds. 
		Then $\lim\limits_{t\to\infty}y_i(t)-y_j(t)\in[z_{\min},z_{\max}]$, where 
		$$z_{\min}=\max_{P_{i,j}}\sum_{k\in P_{i,j}}(-1)^{p_k}I_k^L, \, z_{\max}=\min_{P_{i,j}}\sum_{k\in P_{i,j}}(-1)^{p_k}I_k^R,$$ and $P_{i,j}\in\mathbf{P}_{i,j}$ is a directed path from node $i$ to node $j$, and $p_k=0$, if the original orientation of edge $k$ is consistent with the direction of path $P_{i,j}$, and $p_k=1$ otherwise.
	\end{corollary}	
	\begin{IEEEproof}
		Denote $P_{\alpha}$ as the path such that $z_{\max}=\sum_{k\in P_{\alpha}}(-1)^{p_k}I_k^R$.
		Without loss of generality, suppose the sequencing edges from $i$ to $j$ on path $P_{\alpha}$ are labeled as $1,\ldots, q+1$, where $q\geq 0$, and the sequencing nodes are labeled as $i_0,i_1,\ldots, i_q,i_{q+1}$, where node $i$ corresponds to $i_0$, and node $j$ corresponds to $i_{q+1}$.	
		According to Theorem~\ref{thm_interval}, we have
		$$
		\begin{aligned}
		\lim\limits_{t\to\infty}y_i(t)-y_{i_1}(t)&\leq (-1)^{p_1}I_1^R,\\
		\lim\limits_{t\to\infty}y_{i_1}(t)-y_{i_2}(t)&\leq (-1)^{p_2}I_2^R,\\
		&\vdots\\		
		\lim\limits_{t\to\infty}y_{j}(t)-y_{i_q}(t)&\leq (-1)^{p_{q+1}}I_{q+1}^R.\\
		\end{aligned}		
		$$
		Therefore, $\lim\limits_{t\to\infty}y_i(t)-y_j(t)\leq\sum_{k\in P_{\alpha}}(-1)^{p_k}I_k^R=z_{\max}$.
		The other side of the inequality can be concluded in the same way. 
	\end{IEEEproof}

\begin{figure} [!t]
	\centering
	\subfigure[]
	{\includegraphics[width=.8in]{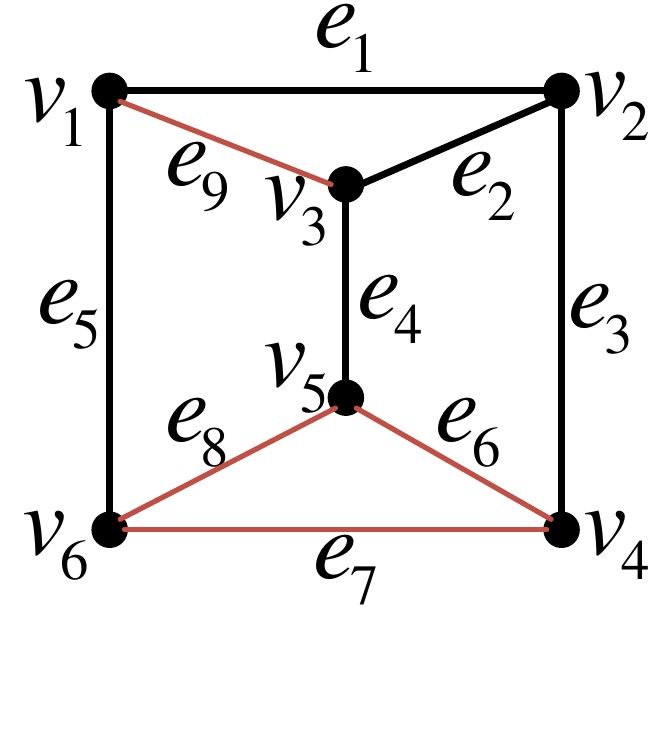}	
		\label{fig_exp1graph}}
	\subfigure[]
		{\includegraphics[width=1.2in]{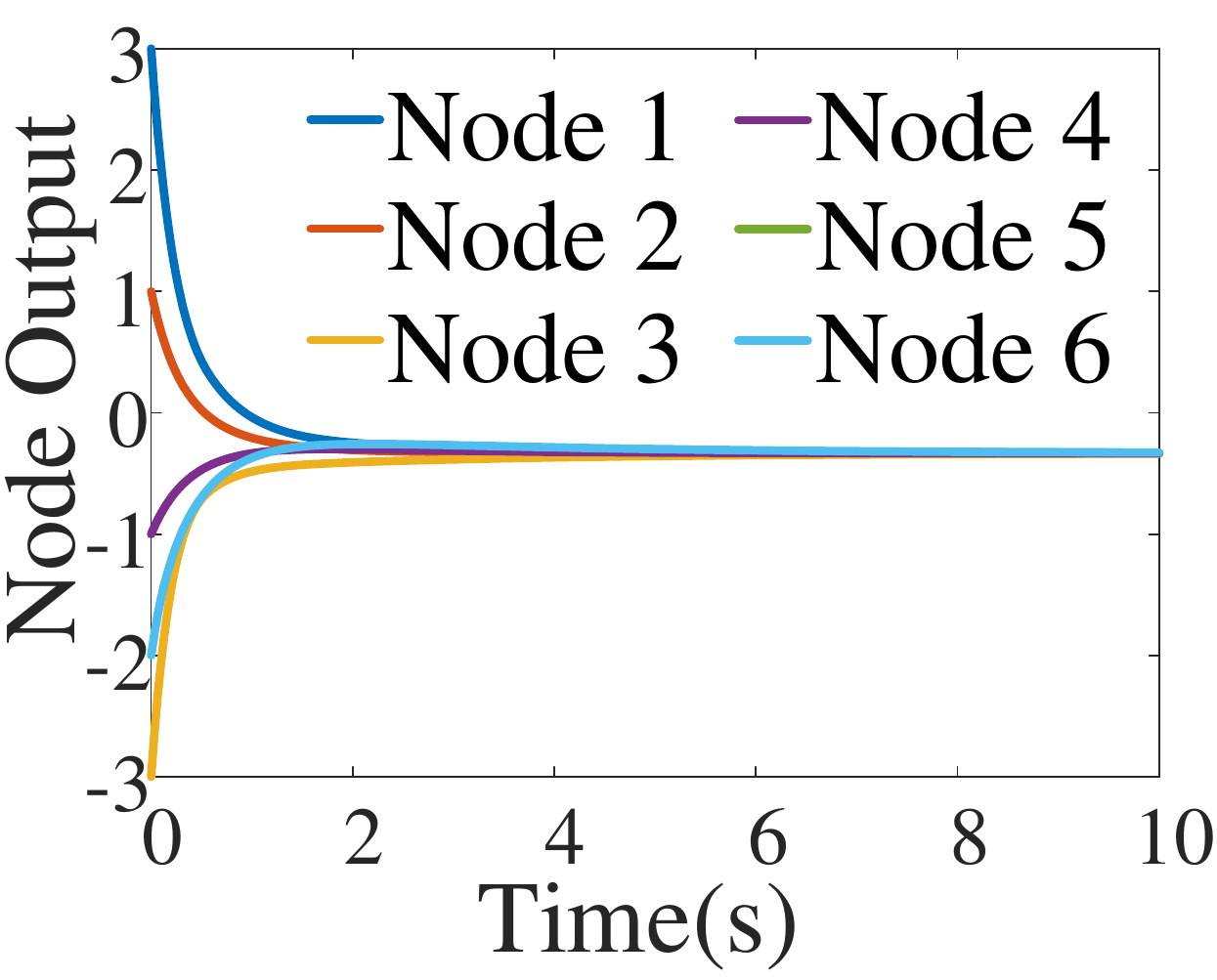}
		 \label{fig_exp1consensus}	}	
	\subfigure[]
	{\includegraphics[width=1.2in]{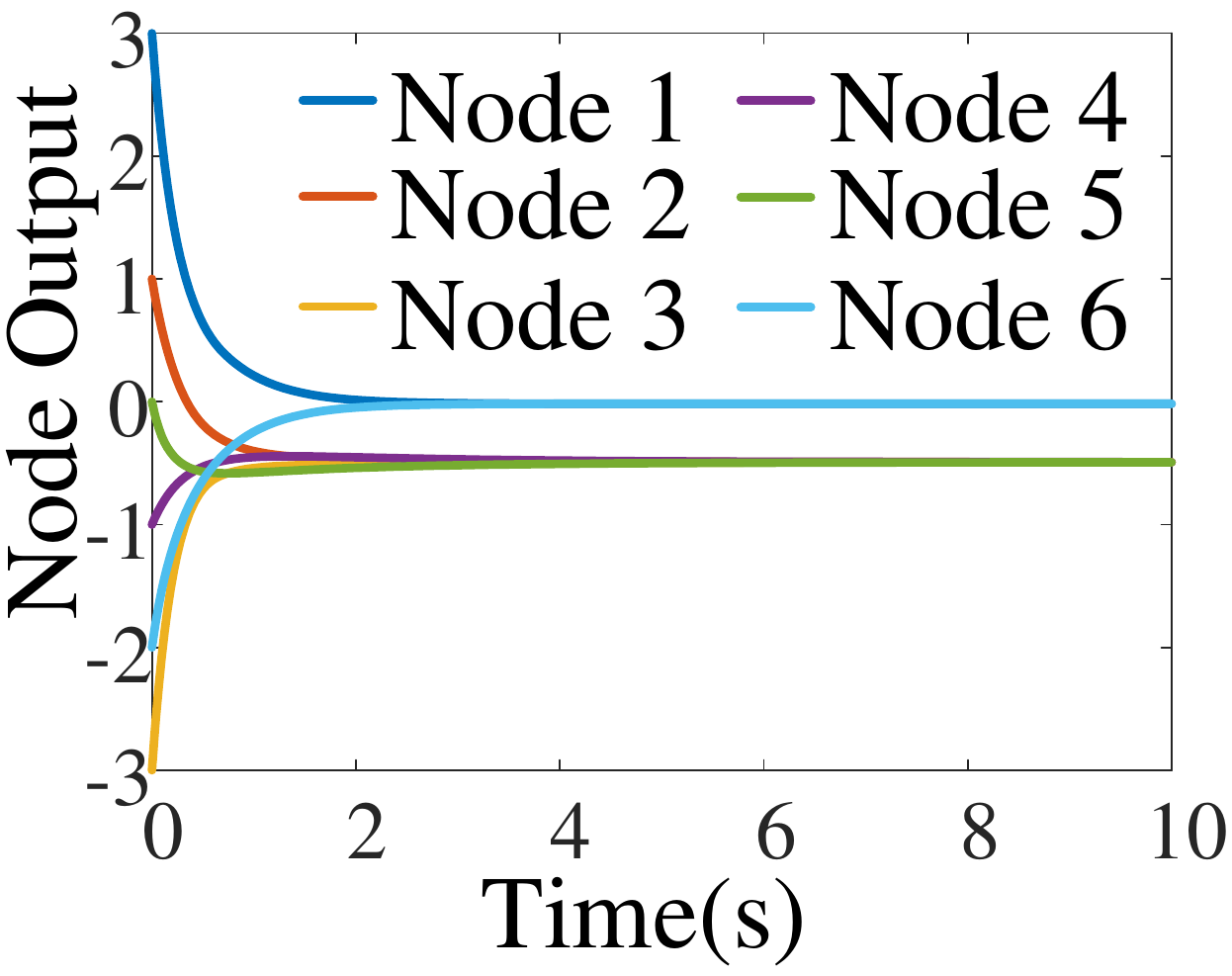}	
		\label{fig_exp1cluster}}		
	\caption{Trajectories of a six-node positive network in Example~\ref{exp_convergence of signedcase}. 
		(a) Underlying graph of the original network, all the nodes and edges $e_k~(k=1,\ldots,5)$ form a spanning tree.
		(b) The agreement case when edges $e_k~(k=1,\ldots,5)$ are strictly positive.
		(c) The clustering case when only edges $e_k~(k=2,\ldots,5)$ are strictly positive.
		}
\end{figure}


\begin{example}\label{exp_convergence of signedcase}
	Consider the network $\mathcal{G}=(\mathcal{V},\mathcal{E}_{\geq})$ shown in Fig.~\ref{fig_exp1graph}, where $\mathcal{V}=\{v_1,\ldots,v_6\}$, and $\mathcal{E}=\{e_1,\ldots,e_9\}$.
	The node dynamics are all single integrators, and therefore Assumption \ref{asp_node} holds.
	%
	%
	%
	If we choose the edge functions for $e_k~(k=1,\ldots,5)$ as
	 \begin{equation}
	 \label{eq_exp1strictly}
	 \mu_k(t)=\zeta_k(t),
	 \end{equation}
	 and the edge functions for $e_k~(k=6,\ldots,9)$ as (see Fig.~\ref{fig_memristor})
	 \begin{equation}
	 \label{eq_exp1notstrictly}
	 \mu_k(t)=\mathrm{sign}(\zeta_k(t))\cdot\max\{|\zeta_k(t)|-1,0\},
	 \end{equation}
	 where $\mathrm{sign}(\cdot)$ is the \emph{signum} function. 
	 Then we have a connected subnetwork spanning all the nodes 
	 and strictly positive edges, i.e., $e_k~(k=1,\ldots,5)$.
	%
	Fig.~\ref{fig_exp1consensus} shows the trajectory of each node with the initial states of the nodes set as $[3,1,-3,-1,0,-2]^T$.
	We can see the outputs of the nodes reach agreement finally, verifying Corollary~\ref{coro_positive+strictly}.
	
	Now we change the edge function of $e_1$ from~\eqref{eq_exp1strictly} to~\eqref{eq_exp1notstrictly}.
	Let $\mathcal{G}_>=(\mathcal{V},\mathcal{E}_>)$ span all nodes and strictly positive edges of $\mathcal{G}$, then $\mathcal{G}_>$ is made up of two connected components, one containing nodes $v_1$ and $v_6$, and one containing nodes $v_2$, $v_3$, $v_4$, and $v_5$.
	Under the same initial condition, the trajectory of each node is depicted in Fig.~\ref{fig_exp1cluster}.
	We can see the outputs form two steady clusters finally, and each cluster corresponds to one connected component in $\mathcal{G}_>$. 
	The distance between the outputs of the two clusters is 0.48, belonging to the interval of equilibria $[-1,1]$, which can be derived from Corollary~\ref{coro_distance}. 
	%
	%
\end{example}
	
Now we have shown that the convergence is always guaranteed for a positive network.
Specifically, if there is a connected subnetwork made up of all nodes and strictly positive edges, then the nodes' outputs will finally reach agreement.
We have also indicated the bounds of the distances between the steady-state outputs of any pair of nodes.
%

\subsection{Signed Networks}
The convergence properties for the signed network of general MEIP systems are quite complicated.
In fact, it is possible that relative outputs $\boldsymbol{\zeta}(t)$ do not converge to a point in $\boldsymbol{I}$ even when all the nodes are in their steady states with the equilibrium input $\boldsymbol{u}(t)=\boldsymbol{0}$. 
Some typical examples are given in~\cite{Zelazo2014,Zelazo2017}, where the networks of single integrators with edge functions represented by~\eqref{eq_linear} form steady clusters, while $\boldsymbol{\zeta}(t)$ does not converge to the equilibria of the corresponding edge functions.
Now we provide a necessary and sufficient condition indicating when the relative outputs $\boldsymbol{\zeta}(t)$ converge to a point in $\boldsymbol{I}$ under the condition that $\lim\limits_{t\rightarrow\infty}\boldsymbol{u}(t)=\mathbf{0}$.
%
%

\begin{proposition}
	\label{thm_signed}
	Consider a signed network system $(\mathcal{G},\Sigma,\Pi)$ with connected graph $\mathcal{G}=(\mathcal{V},\mathcal{E})$ represented by~\eqref{eq_plant}-\eqref{eq_umu}.
	%
	Suppose Assumption~\ref{asp_node} holds and $\lim\limits_{t\rightarrow\infty}\boldsymbol{u}(t)=\mathbf{0}$.
	%
	Then $\lim\limits_{t\rightarrow\infty}\boldsymbol{\zeta}(t)=\tilde{\boldsymbol{\upzeta}}$ exists and $\tilde{\boldsymbol{\upzeta}}\in\boldsymbol{I}$, if and only if for any $k\notin\mathcal{E}_{\geq}$,  $\lim\limits_{t\rightarrow\infty}\zeta_k(t)=\tilde{\upzeta}_k$ exists and $\tilde{\upzeta}_k\in I_k$.	
\end{proposition}
\begin{IEEEproof}
	(Sufficiency) 
	With the precondition $\lim\limits_{t\rightarrow\infty}\boldsymbol{u}(t)=\mathbf{0}$, we have $\lim\limits_{t\rightarrow\infty}\boldsymbol{y}(t)=\tilde{\mathbf{y}}\in\boldsymbol{\sigma}(\mathbf{0})$, therefore $\lim\limits_{t\rightarrow\infty}\boldsymbol{\zeta}(t)=\tilde{\boldsymbol{\upzeta}}=E^T\tilde{\mathbf{y}}$ exists.
	Assume $\lim\limits_{t\rightarrow\infty}\boldsymbol{\zeta}(t)\notin \boldsymbol{I}$, then there must exist a set $\tilde{\mathcal{E}}\subset \mathcal{E}_{\geq}$ such that $\lim\limits_{t\rightarrow\infty}\zeta_k(t)\notin I_k$ if $k\in \tilde{\mathcal{E}}$. 
	Denote $\tilde{\mathcal{V}} \subset \mathcal{V}$ as the set of nodes which are incident to at least one edge in $\tilde{\mathcal{E}}$.
	Suppose node $p$ has the maximum equilibrium output among all the nodes in $\tilde{\mathcal{V}}$, i.e., $\lim\limits_{t\rightarrow\infty}y_p(t) = \max_{i\in \tilde{\mathcal{V}}} \lim\limits_{t\rightarrow\infty}y_i(t)$. 
	Note that $\forall~k\notin \tilde{\mathcal{E}}$, $\lim\limits_{t\rightarrow\infty}\mu_k(t)=0$. 
	As a result, $\lim\limits_{t\rightarrow\infty}u_p(t)<0$, which contradicts the precondition that $\lim\limits_{t\rightarrow\infty}\boldsymbol{u}(t)=\mathbf{0}$. Thus $\tilde{\mathcal{E}}$ does not exist, i.e., $\lim\limits_{t\rightarrow\infty}\zeta_k(t)\in I_k,\forall~k\in\mathcal{E}_{\geq}$. Therefore, $\lim\limits_{t\rightarrow\infty}\boldsymbol{\zeta}(t)=\tilde{\boldsymbol{\upzeta}}\in\boldsymbol{I}$.
	
	(Necessity) This is straightforward and omitted.
\end{IEEEproof}

By using the same technique as in the proof of Proposition~\ref{thm_signed}, we can conclude the following corollary for the consensus case, i.e., $\lim\limits_{t\rightarrow\infty}\boldsymbol{\zeta}(t)=\mathbf{0}$.
\begin{corollary}
	\label{coro_signedstrictly}
	Consider a signed network system $(\mathcal{G},\Sigma,\Pi)$ with connected graph $\mathcal{G}=(\mathcal{V},\mathcal{E})$ represented by~\eqref{eq_plant}-\eqref{eq_umu}.
	%
	Suppose Assumption~\ref{asp_node} holds and $\lim\limits_{t\rightarrow\infty}\boldsymbol{u}(t)=\mathbf{0}$.
	Then 
	%
	$\lim\limits_{t\rightarrow\infty}\boldsymbol{\zeta}(t)=\mathbf{0}$, if and only if for any $k\notin\mathcal{E}_>$, $\lim\limits_{t\rightarrow\infty}\zeta_k=0$.
\end{corollary}

Proposition~\ref{thm_signed} and Corollary~\ref{coro_signedstrictly} show that, when there are non-positive (non-strictly positive) edges added to the positive (strictly positive) network,
$\boldsymbol{\zeta}(t)$ can still converge to the equilibria of the corresponding edge functions, 
if relative outputs of the nodes connected by the non-positive (non-strictly positive) edges converge to their equilibria.
However, one may note that with Proposition~\ref{thm_signed} and Corollary~\ref{coro_signedstrictly}, before executing the interaction protocol, we still cannot decide whether 
$\lim\limits_{t\rightarrow\infty}\boldsymbol{u}(t)=\mathbf{0}$, nor decide whether those non-positive (non-strictly positive) edges can converge to their equilibria.
We will further discuss these conditions in Section~\ref{sec:single}.
Before we proceed, we provide the following proposition showing an important property of the nodes only incident to strictly positive edges.

\begin{proposition}
	\label{pro_clustermax}
	Consider a signed network system $(\mathcal{G},\Sigma,\Pi)$ with connected graph $\mathcal{G}=(\mathcal{V},\mathcal{E})$ represented by~\eqref{eq_plant}-\eqref{eq_umu}. Suppose Assumption~\ref{asp_node} holds and $\lim\limits_{t\rightarrow\infty}\boldsymbol{u}(t)=\mathbf{0}$. Then the following inequality holds,
	\begin{equation}
	\label{eq_minmax}
	\min_{i\in \mathtt{N}_p}\lim\limits_{t\to\infty}y_i(t)\leq \lim\limits_{t\to\infty}y_p(t)\leq\max_{i\in \mathtt{N}_p}\lim\limits_{t\to\infty}y_i(t),
	\end{equation}
	 where $p$ is any node only incident to strictly positive edges.
	Equalities in~\eqref{eq_minmax} hold at the same time, i.e., $\min_{i\in \mathtt{N}_p}\lim\limits_{t\to\infty}y_i(t)= \lim\limits_{t\to\infty}y_p(t)$ if and only if $\lim\limits_{t\to\infty}y_p(t)=\max_{i\in \mathtt{N}_p}\lim\limits_{t\to\infty}y_i(t)$.
\end{proposition}
\begin{IEEEproof}
	Suppose when $t\to\infty$, $y_p(t)>\max_{i\in \mathtt{N}_p}y_i(t)$, where node $p$ is only incident to strictly positive edges, then $\lim\limits_{t\to\infty}u_p(t)<0$, which contradicts the precondition $\lim\limits_{t\to\infty}\boldsymbol{u}(t)=\mathbf{0}$.
	Therefore, we can get $\lim\limits_{t\to\infty}y_p(t)\leq\max_{i\in \mathtt{N}_p}\lim\limits_{t\to\infty}y_i(t)$.
	In the same way, We can conclude $\lim\limits_{t\to\infty}y_p(t)\geq\min_{i\in \mathtt{N}_p}\lim\limits_{t\to\infty}y_i(t)$ .
	
	Now if $\lim\limits_{t\to\infty}y_p(t)=\min_{i\in \mathtt{N}_p}\lim\limits_{t\to\infty}y_i(t)$, in order to make $\lim\limits_{t\to\infty}u_p(t)=0$, it requires that $\lim\limits_{t\to\infty}y_i(t)=\lim\limits_{t\to\infty}y_p(t),~\forall~i\in \mathtt{N}_p$, as a result, 
	$\lim\limits_{t\to\infty}y_p(t)=\max_{i\in \mathtt{N}_p}\lim\limits_{t\to\infty}y_i(t)$.
	We can conclude in the other way around with the same method.
\end{IEEEproof}

Proposition~\ref{pro_clustermax} shows that, when $\lim\limits_{t\to\infty}\boldsymbol{u}(t)=\mathbf{0}$, the nodes only incident to strictly positive edges, cannot be the only one with the maximum or minimum steady output among all the nodes.
%

%

\section{Equivalent Circuit Models and\\ Signed Nonlinear Networks\label{sec:single}}
%
In this section, we further investigate the convergence properties of signed nonlinear networks by exploring  connections with circuit theory.  In particular, we examine the role of \emph{equivalent circuits} in the analysis of these networks when the node dynamics are represented by nonlinear integrators.  

\subsection{Circuit Interpretations}

\begin{figure}[!htb]
	\centering
	\subfigure[]{\includegraphics[height=2.5cm]{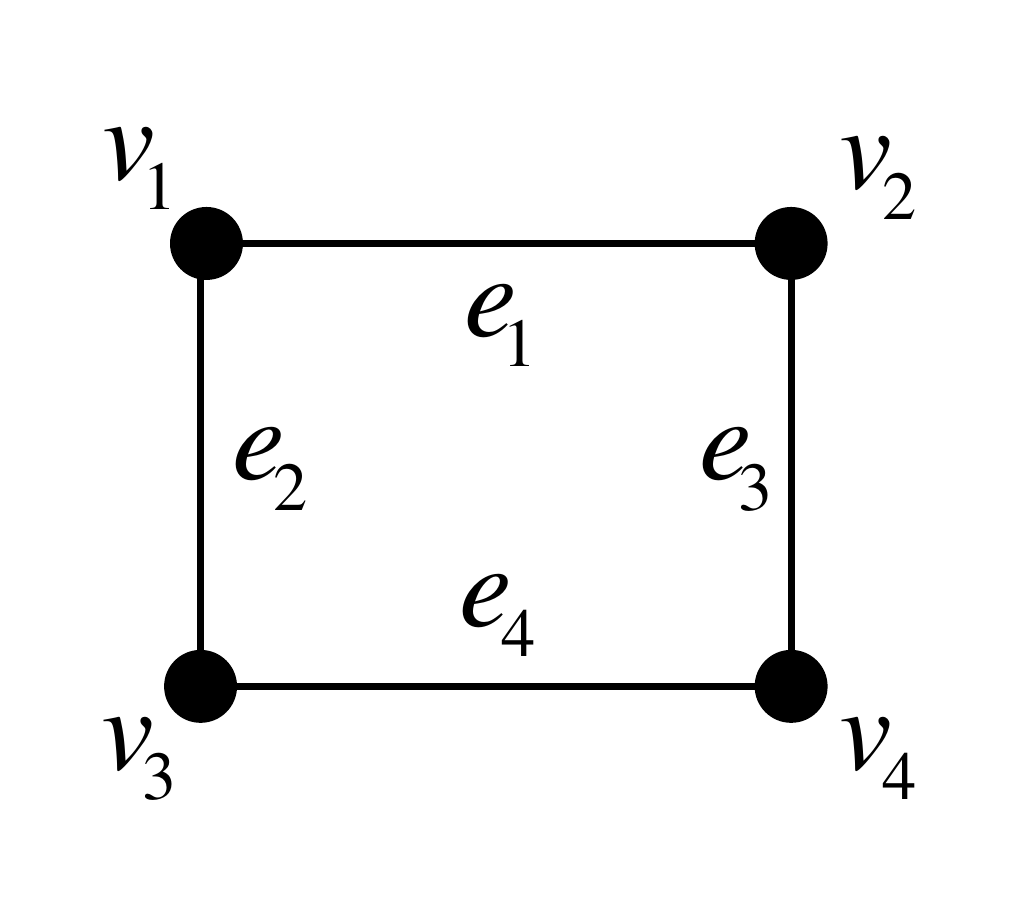}\label{fig_circuit_graph}}
	\hspace{0.2in}
	\subfigure[]{\includegraphics[height=2.5cm]{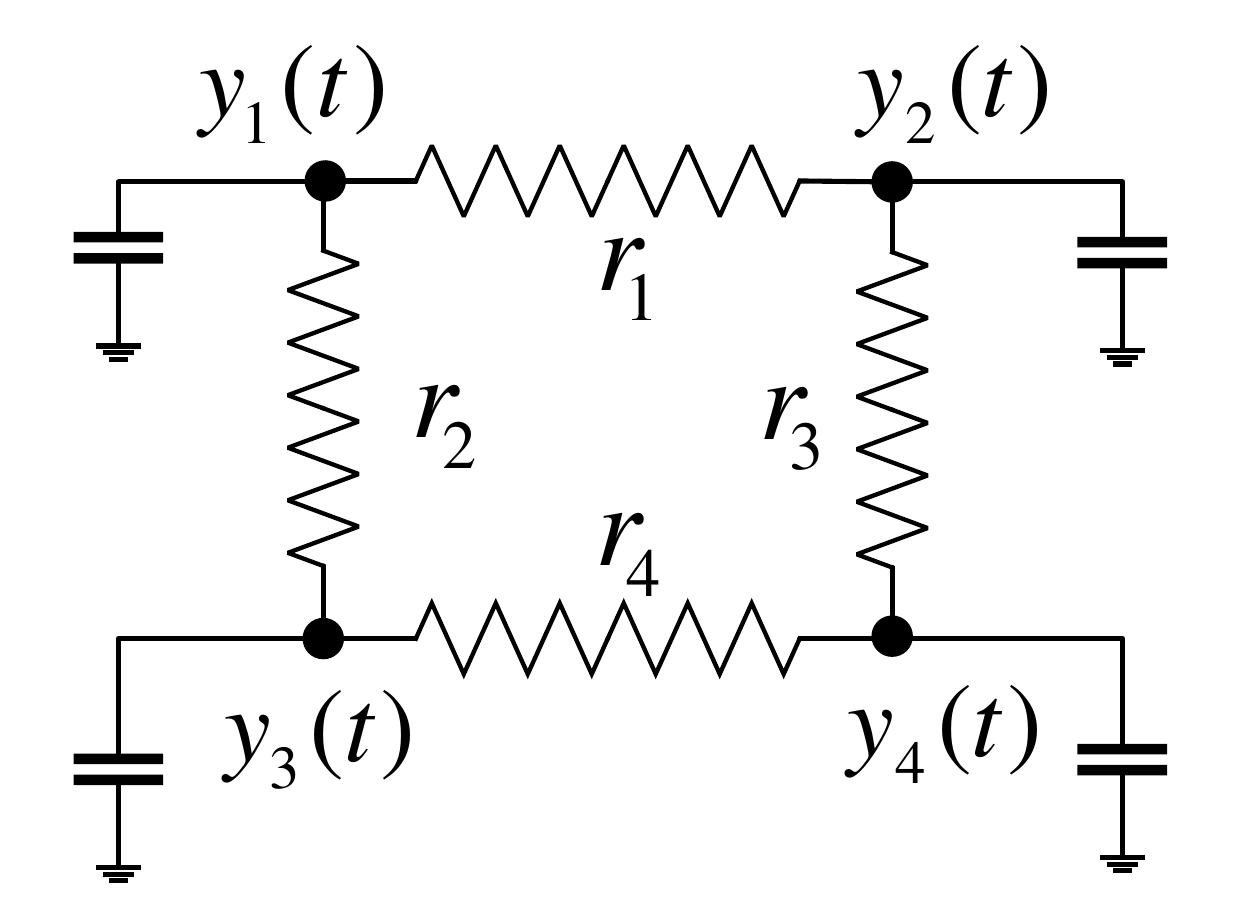}\label{fig_circuit_circuit}}
	\caption{Circuit interpretation of a network of single integrators. (a) Underlying graph of a four-node network and (b) the corresponding circuit.}
	\label{fig_circuit}
\end{figure}

As we have mentioned in Remark~\ref{rmk_circuit}, the network model formulated in Section~\ref{sec:networkmodel} has a circuit interpretation.
For example, the network system comprised of single integrators can be interpreted as a resistor-capacitor (RC) circuit.
Fig.~\ref{fig_circuit_graph} shows the underlying graph of a four-node network, and Fig.~\ref{fig_circuit_circuit} shows the corresponding circuit interpretation. 
In Fig.~\ref{fig_circuit_circuit}, the capacitance of each capacitor is 1 Farad.
As we have mentioned in Remark~\ref{rmk_circuit}, $\boldsymbol{y}(t)$ represents the \emph{potentials} of the nodes, that is, $y_i(t)$ stands for the potential of node $v_i$ with respect to the ground, $i=1,\ldots,4$.
The edge $e_k$ in Fig.~\ref{fig_circuit_graph} corresponds to the \emph{voltage-controlled} resistor $r_k$, $k=1,\ldots,4$.
The \emph{potential difference} between two ends of an edge is called a \emph{tension}, which is the \emph{voltage drop} on the resistor, and its stacked form corresponds to $\boldsymbol{\zeta}(t)$ in our network model. 
The tension will result in \emph{flow}, i.e., the \emph{current} on the resistor in the circuit, corresponding to $\boldsymbol{\mu}(t)$.
The flow $\mu_k(t)$ on edge $k$ is generated according to tension $\zeta_k(t)$ as well as the edge functions defined in~\eqref{eq_umu}. 
In circuit theory, such edge functions are exactly the \emph{current-voltage 
	functions} 
of the voltage-controlled resistors.

According to Definition~\ref{def_signed}, if edge $k$ is positive, then the inner product $\mu_k(t)\zeta_k(t)\geq0$, meaning the corresponding resistor $r_k$, is generally \emph{energy consuming}, and it does not consume energy (nor produce energy) if and only if $\mu_k(t)=0$.
Moreover, for a strictly positive edge $k$, the resistor always consumes energy unless $\mu_k(t)=\zeta_k(t)=0$.
On the contrary, a negative edge means its corresponding ``resistor'' can produce energy when $\mu_k(t)\zeta_k(t)<0$. Practically, this can only be realized if it contains sources.

Still, from the energy perspective, $(u_i(t)-\mathrm{u}_i)(y_i(t)-\mathrm{y}_i)$ is the energy that flows into node $i$ compared with the steady-state equilibrium I/O pair $(\mathrm{u}_i,\mathrm{y}_i)$, while $S_i(x_i(t))$ is the energy stored in node $i$. 
For the MEIP plant, since $\dot{S}_i(x_i(t))\leq(u_i(t)-\mathrm{u}_i)(y_i(t)-\mathrm{y}_i)$, the increased energy storage is no more than the energy that flows into it.
In terms of the single integrator, $S_i(x_i(t)):=\frac{1}{2}(x_i(t)-\mathrm{y}_i)^2$, and $\dot{S}_i(x_i(t))=(u_i(t)-\mathrm{u}_i)(y_i(t)-\mathrm{y}_i)$, meaning the increased energy storage is exactly the energy that flows into it.

\subsection{Equivalent Edge Functions\label{sec:equivalent}}


\emph{Effective resistance} is often used as a distance metric on networks~\cite{Klein1993}.
If all the edge functions in the network are in the form of~\eqref{eq_linear}, we obtain $\boldsymbol{u}=-EWE^T\boldsymbol{y}$, where $W=\mathbf{diag}\{w_1,\ldots,w_{|\mathcal{E}|}\}$ is the diagonal edge weight matrix. 
In graph theory, $L(\mathcal{G}):=EWE^T$ is the weighted Laplacian matrix. 
The effective resistance between nodes $p$ and $q$, denoted as $\bar{r}_{pq}$, can be calculated as~\cite{Klein1993}:
$$
\begin{aligned}
\bar{r}_{pq}&=(\mathbf{e}_p-\mathbf{e}_q)^TL(\mathcal{G})^{\dagger}(\mathbf{e}_p-\mathbf{e}_q), 
\end{aligned}
$$
where $L(\mathcal{G})^{\dagger}$ is the Moore-Penrose pseudoinverse of the weighted Laplacian matrix, $\mathbf{e}_i\in\mathbb{R}^{|\mathcal{V}|}$ is the $i$-th basis of $\mathbb{R}^{|\mathcal{V}|}$, that is, $[\mathbf{e}_i]_i=1$, and $[\mathbf{e}_i]_j=0$ if $j\neq i$.

Effective resistance has a clear circuit interpretation.
When $\boldsymbol{u}=\mathbf{0}$, it means all the nodes are in their steady states and the network corresponds to a \emph{resistive} circuit.\footnote{In circuit theory, a resistive circuit is a circuit containing only resistors and sources, without dynamic elements such as inductors or capacitors~\cite{chua1969introduction}. In our cases, though there are capacitors in the corresponding circuit in Fig.~\ref{fig_circuit_circuit}, since $\boldsymbol{u}=\mathbf{0}$, the nodes' states will not change, making the property of the circuit similar to that of a resistive circuit.}
If all the edge functions are in the form of~\eqref{eq_linear}, then each corresponding resistor in the circuit is with constant resistance $r_k=\frac{1}{w_k}$.
Take nodes $p,q\in\mathcal{V}$ as two terminals of interest. When we add a voltage source with the voltage value $\zeta_{pq}$ outside the two terminals, the current flow into the two-terminal network is $\zeta_{pq}/\bar{r}_{pq}$, where $\bar{r}_{pq}$ is exactly the effective resistance of the original network.
In this case, the resistors of the resistive circuit between terminals $p$ and $q$ can be replaced by a single resistor whose resistance equals to $\bar{r}_{pq}$.

Now we generalize the concept of \emph{effective resistance} for nonlinear networks, and introduce the notion of \emph{equivalent edge functions}. 
%
As with effective resistance, we first identify two nodes $p,q \in \mathcal{V}$ to represent the terminals of interest in the network. 
	%
	Consider now the addition of a virtual edge $\bar{k}$ connecting nodes $p$ and $q$, 
	and define an augmented graph $\bar{\mathcal{G}}=(\mathcal{V},\bar{\mathcal{E}})$, where $\bar{\mathcal{E}}=\mathcal{E}\cup \{\bar{k}\}$, with its incidence matrix denoted as $\bar{E}$. 
	The stacked tension and flow of the augmented network are denoted as $\bar{\boldsymbol{\zeta}}$ and $\bar{\boldsymbol{\mu}}$, respectively. The corresponding network equations for the augmented graph are
	\begin{equation}
	\label{augnet_equations}
	\bar{E}^T\boldsymbol{y}=\bar{\boldsymbol{\zeta}}, \;\boldsymbol{\mu}=\Psi(\boldsymbol{\zeta}), \;\bar{E}\bar{\boldsymbol{\mu}}=\mathbf{0}.
	\end{equation}
Equivalent edge functions are now determined by the solutions of these equations, which we formalize below.
\begin{definition}
Consider the augmented network system $(\mathcal{\bar{G}},\Sigma,\Pi)$ with virtual edge $\bar{k}=(p,q)$.  For each given $\zeta_{\bar{k}}$, if there exists a unique $(\bar{\boldsymbol{\zeta}},\bar{\boldsymbol{\mu}})$ and some $\boldsymbol{y}\in\mathbb{R}^{|\mathcal{V}|}$ to the network equations \eqref{augnet_equations}, then the flow $\mu_{\bar{k}}$ on the virtual edge $\bar{k}$ can be represented as a function of $\zeta_{\bar{k}}$, 
	which we denote as $\mu_{\bar{k}}=-\bar{\psi}_{pq}(\zeta_{\bar{k}})$. We term $\bar{\psi}_{pq}(\cdot):\mathbb{R}\rightarrow\mathbb{R}$ the \emph{equivalent edge function} between nodes $p$ and $q$ in the original network system $(\mathcal{G}, \Sigma, \Pi)$.
\end{definition}

\begin{remark}
Note that for the augmented network, $\boldsymbol{u}=-\bar{E}\bar{\boldsymbol{\mu}}=\mathbf{0}$. Thus the augmented network corresponds to a \emph{resistive circuit}.
\end{remark}
	%


The concept of equivalent edge functions is well suited for the circuit interpretation of the effective resistance.
We use the following example to detail it.

\begin{figure}[!htb]
	\centering
	\subfigure[]{\includegraphics[width=3.5cm]{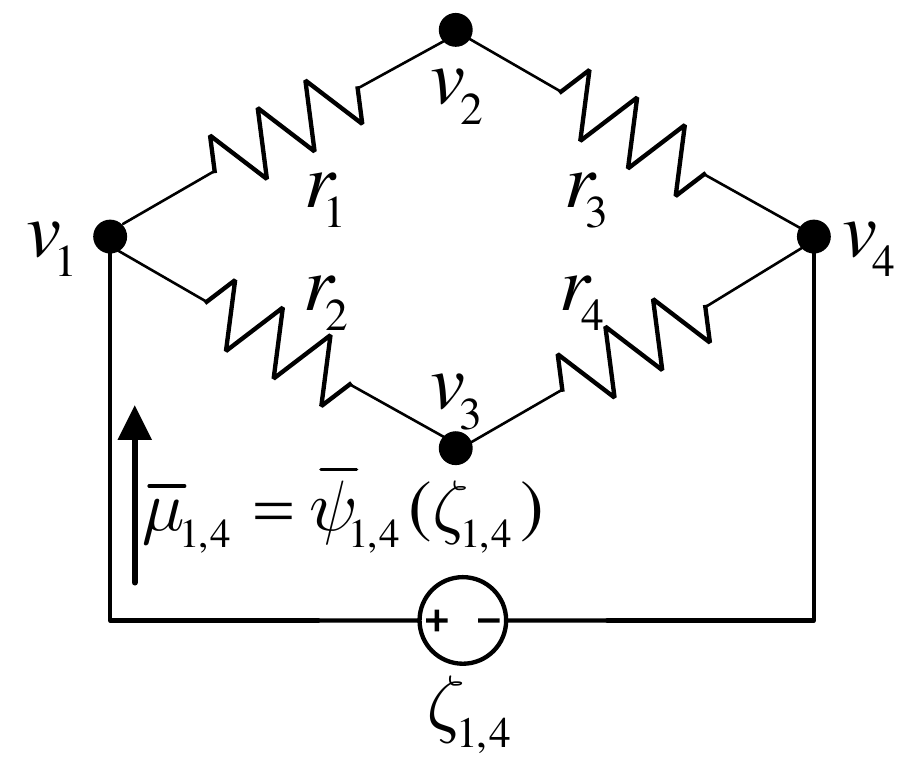}\label{fig_fourresistors}}
	\hspace{0.1in}
	\subfigure[]{\includegraphics[width=3.1cm]{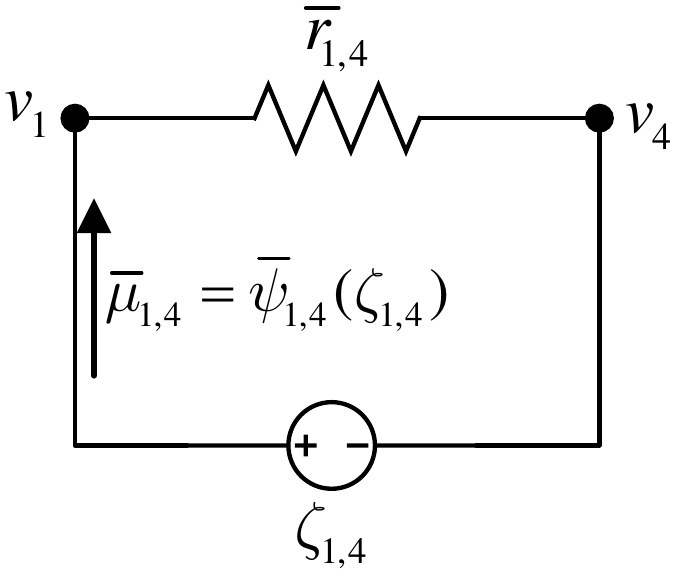}\label{fig_equivalentsub}}
	\caption{Equivalent resistor of a two-terminal circuit network, with a voltage source connecting the two terminals. (a) Original circuit and (b) Equivalent resistor of the two-terminal circuit.}
	\label{fig_equivalent}
\end{figure}

\begin{example}
As shown in Fig.~\ref{fig_equivalent},
consider a voltage drop $\zeta_{1,4}$ from $v_1$ to $v_4$ caused by a voltage source.
%
%
If for each value of $\zeta_{1,4}$, there is a unique current flowing from $v_1$ to $v_4$ in the two-terminal circuit system, then the four resistors in Fig.~\ref{fig_fourresistors} can be simplified as one \emph{equivalent} resistor $\bar{r}_{1,4}$, as shown in Fig.~\ref{fig_equivalentsub}.
In this case, the voltage source can be regarded as the newly-added virtual edge $\bar{k}$,
and the resistance of the equivalent resistor $\bar{r}_{1,4}$ may not be a constant, but its 
 equivalent current-voltage function corresponds to the equivalent edge function between $v_1$ and $v_4$ in the original network system.
\end{example}

\begin{remark}	
	Although the equivalent edge function $\bar{\psi}_{pq}(\cdot)$ takes a similar form as in~\eqref{eq_edge}, they have different meanings. The equivalent edge function corresponds to the equivalent current-voltage function of the equivalent resistor, while the edge function in the form of~\eqref{eq_edge} corresponds to the current-voltage function of an actual resistor in the circuit. 
	Stated in another way, edge function shown in~\eqref{eq_edge} models a real interaction between neighboring agents, while the equivalent edge function is meant to replace the entire network interaction between two nodes by a single, virtual, edge function.
\end{remark}


Now we discuss the existence of the equivalent edge functions.  We first need the following result from circuit theory. 
\begin{lemma}[\cite{chua1969introduction}]
	\label{lm_circuitunique}
	Consider a circuit containing only resistors and independent voltage sources. 
	If the current-voltage function of each resistor is strictly monotonically increasing with the current tending to $\pm\infty$ as the voltage tends to $\pm\infty$, and if the circuit contains no cycles of voltage sources, then for each pair of the voltage source values, the current and voltage drop on each resistor, as well as the current on each source, are unique.
\end{lemma}

We now have the following proposition for the existence of the equivalent edge function.

\begin{proposition}
	\label{pro_existence}
	Consider a strictly positive network system $(\mathcal{G},\Sigma,\Pi)$ with connected graph  $\mathcal{G}=(\mathcal{V},\mathcal{E}_>)$ represented by~\eqref{eq_plant}-\eqref{eq_umu}.
	Identify two nodes $p,q\in\mathcal{V}$ as the terminals of interest.
	If {for each} $k\in\mathcal{E}_>$, its edge function $\mu_k(t)=\psi_k(\zeta_k(t))$ is strictly monotonically increasing, and $\mu_k(t)\to\pm\infty$ as $\zeta_k(t)\to\pm\infty$, then the equivalent edge function of the two-terminal network between nodes $p$ and $q$ exists.
\end{proposition}
\begin{IEEEproof}
	We add a virtual edge $\bar{k}$ which connects nodes $p$ and $q$.
	When the value of $\zeta_{\bar{k}}(t)$ is given, its corresponding circuit is equivalent to adding a voltage source outside of the two-terminal circuit system, and the voltage value is $\zeta_{\bar{k}}(t)$, as shown in Fig.~\ref{fig_fourresistors}.
	Since there is only one voltage source, according to Lemma~\ref{lm_circuitunique}, when $\boldsymbol{u}(t)=\mathbf{0}$, the flow $\mu_{\bar{k}}$ on the virtual edge $\bar{k}$ is unique.
	By varying the value of $\zeta_{\bar{k}}(t)$ from $-\infty$ to $+\infty$, we can obtain the equivalent edge function of the two-terminal network between nodes $p$ and $q$.
	%
\end{IEEEproof}
%
%

\begin{remark}
When all the edge functions are in the form~\eqref{eq_linear} with positive scalar weights, the equivalent edge function between any two nodes $p,q\in\mathcal{V}$ always exists, and is $\bar{\mu}_{pq}(t)=\zeta_{pq}(t)/\bar{r}_{pq}$, where $\bar{r}_{pq}$ is the effective resistance between $p$ and $q$, verifying Proposition~\ref{pro_existence}. 
\end{remark}

	Generally, it is difficult to obtain an analytical characterization of the equivalent edge function when the edge functions in the network are nonlinear.
	Here we present an algorithm to obtain an approximate result, as shown in Algorithm~\ref{alg}. 
	%
	First, we obtain the resistive circuit corresponding to the network system $(\mathcal{G},\Sigma,\Pi)$, and add a voltage source between terminals $p$ and $q$ (in Line~\ref{ln_obtaincircuit}). We obtain a finite set of potential differences $\{\zeta_{pq}\}$ for the two terminals by sampling in the interval $[-N,N]$, where $N$ is an arbitrary large positive number (in Line~\ref{ln_specify}). The interval defines the set of interest, and the approximated equivalent edge function's accuracy depends on its resolution.
	Then, for each $\zeta_{pq}$ in the interval, we set it as the value of the voltage source (in Line~\ref{ln_setvoltage}), and calculate the flow into the two-terminal network,\footnote{In circuit theory, this process is the calculation of the operating point of the circuit.} while satisfying KVL and KCL, i.e., equation~\eqref{eq_zetay} and~\eqref{eq_umu} (in Line~\ref{ln_calculate}).
	%
	The equivalent edge function is finally approximated by using interpolation techniques in Line~\ref{ln_approximate}.

\begin{algorithm}[!h]
\caption{Computation of Equivalent Edge Functions} 
\label{alg}	
\begin{algorithmic}[1]
\State Obtain the corresponding resistive circuit, and add a voltage source between $p$ and $q$; \label{ln_obtaincircuit}
\State Obtain a finite set $\{\zeta_{pq}\}$ by sampling the interval $[-N,N]$; \label{ln_specify}
\vspace{-12pt}
\ForAll {$\zeta_{pq}$}
\State {Set $\zeta_{pq}$ as the value of voltage source; \label{ln_setvoltage}
}
\State Calculate the flow on the voltage source $\bar{\mu}_{pq}$ while satisfying KVL and KCL; \label{ln_calculate}
\EndFor
\State Approximate the equivalent edge function by interpolation based on $\{(\zeta_{pq},\bar{\mu}_{pq})\}.$ \label{ln_approximate}
\end{algorithmic}
\end{algorithm}	


\subsection{Cocontent Function}
\begin{figure} [!htb]
	\centering
	\includegraphics[height=1.5in]{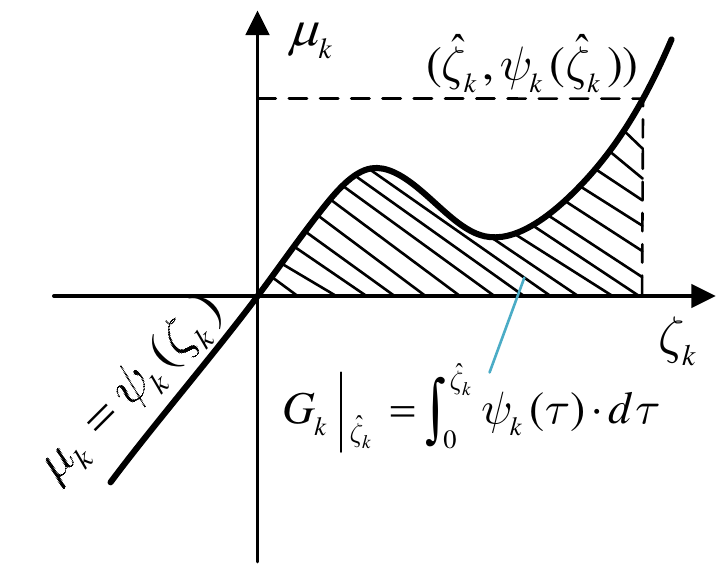}
	\caption{Illustration of the cocontent function and its relationship with the edge function.}
	\label{fig_cocontent}
\end{figure}

Another important concept that we borrow from circuit theory is the cocontent function~\cite{Parodi2018}.
The cocontent function of an edge $k$ when its tension $\zeta_k$ is specified, is defined as 
$$
G_k\lvert_{\zeta_k}=\int_{0}^{\zeta_k}\psi_{k}(\tau)\cdot d\tau.
$$
The relationship between the edge function $\mu_k=\psi_k(\zeta_k)$ and its cocontent function can be represented in the $(\zeta_k,\mu_k)$ plane as shown in Fig.~\ref{fig_cocontent}.
It is easy to verify that the cocontent of a positive edge is always greater than zero. 
%
Besides, the cocontent of a strictly positive edge becomes zero if and only if $\zeta_k=0$.	
Correspondingly, we can define the cocontent of a two-terminal network as the sum of the cocontent of all the edges, which is denoted as $\mathbf{G}=\sum_{k=1}^{|\mathcal{E}|}G_k\lvert_{\zeta_k}$. 

For a strictly positive network system $(\mathcal{G},\Sigma,\Pi)$ whose edge functions are all monotonically increasing, we define an augmented network by adding a virtual edge $\bar{k}$ connecting two nodes of interest.
The stacked tension and flow of the augmented network are denoted as $\bar{\boldsymbol{\zeta}}$ and $\bar{\boldsymbol{\mu}}$, respectively, and the incidence matrix denoted as $\bar{E}$.
We show in Lemma~\ref{lm_cocontent} that the cocontent $\sum_{k=1}^{|\mathcal{E}|}G_k\lvert_{\zeta_k}$ reaches the minimum when $\bar{E}\bar{\boldsymbol{\mu}}=\mathbf{0}$.
%

\begin{lemma}
	\label{lm_cocontent}
	Consider a strictly positive network system $(\mathcal{G},\Sigma,\Pi)$ with connected graph $\mathcal{G}=(\mathcal{V},\mathcal{E}_>)$ represented by~\eqref{eq_plant}-\eqref{eq_umu}, and the augmented graph $\bar{\mathcal{G}}=(\mathcal{V},\bar{\mathcal{E}})$ obtained by adding the virtual edge $\bar{k}$ (with $\bar{\mathcal{E}}=\mathcal{E}_>\cup \{\bar{k}\}$).
	%
	Suppose for each $k\in\mathcal{E}_>$, its edge function $\mu_k=\psi_k(\zeta_k)$ is monotonically increasing.
	For any fixed $\zeta_{\bar{k}}$, if there exists $\bar{\boldsymbol{\zeta}}^0,\bar{\boldsymbol{\mu}}^0,\boldsymbol{y}^0$, 
	such that $\bar{E}^T\boldsymbol{y}^0=\bar{\boldsymbol{\zeta}}^0$, 
	$\boldsymbol{\mu}^0=\Psi(\boldsymbol{\zeta}^0)$ and $\bar{E}\bar{\boldsymbol{\mu}}^0=\mathbf{0}$, then the cocontent $\sum_{k=1}^{|\mathcal{E}|}G_k\lvert_{\zeta_k}$ of the network system $(\mathcal{G},\Sigma,\Pi)$ reaches its minimum at $(\bar{\boldsymbol{\zeta}}^0,\bar{\boldsymbol{\mu}}^0,\boldsymbol{y}^0)$.
\end{lemma}
\begin{IEEEproof}
	%
	We take $
	\bar{\boldsymbol{\zeta}}^0:=(\zeta^0_1,\ldots,\zeta^0_{|\mathcal{E}|},\zeta^0_{\bar{k}})^T
	$ and $
	\bar{\boldsymbol{\mu}}^0:=(\mu^0_1,\ldots,\mu^0_{|\mathcal{E}|},\mu^0_{\bar{k}})^T
	$, corresponding to the case when $\zeta_{\bar{k}}=\zeta^0_{\bar{k}}$. 
	Since all edge functions $\mu_k=\psi_k(\zeta_k)$, $k\in\mathcal{E}_>$ are by assumption monotonically increasing, the inequality
	\begin{equation}
	\label{eq_monotone}
	(\mu_k-\mu_k^0)(\zeta_k-\zeta_k^0)\geq0
	\end{equation}
	holds. Choose $\zeta_k=\zeta_k^0+d\tau$, $\mu_k=\psi_k(\zeta_k)$, then inequality~\eqref{eq_monotone} becomes
	$$
	[\psi_k(\zeta_k)-\mu_k^0]\underbrace{(\zeta_k-\zeta_k^0)}_{d\tau}\geq0.
	$$
	We now integrate with respect to the tension, taking $\zeta_k^0$ as the lower limit and a generic $\zeta_k$ as the upper limit to obtain
	%
	$$
	\int_{\zeta_k^0}^{\zeta_k}[\psi_k(\tau)-\mu_k^0]d\tau\geq0
	$$
	$$
	\Rightarrow~~\underbrace{\int_{\zeta_k^0}^{\zeta_k}\psi_k(\tau)d\tau}_{G_k\lvert_{\zeta_k}-G_k\lvert_{\zeta_k^0}}\geq\int_{\zeta_k^0}^{\zeta_k}\mu_k^0d\tau=(\zeta_k-\zeta_k^0)\cdot\mu_k^0,
	$$
	that is,
	$$
	G_k\lvert_{\zeta_k}-G_k\lvert_{\zeta_k^0}\geq (\zeta_k-\zeta_k^0)\cdot\mu_k^0.
	$$
	By adding all the $|\mathcal{E}|$ inequalities term by term, we get
	\begin{equation}
	\label{eq_cocontentinequality}
	\sum_{k=1}^{|\mathcal{E}|}G_k\lvert_{\zeta_k}-\sum_{k=1}^{|\mathcal{E}|}G_k\lvert_{\zeta_k^0}\geq\sum_{k=1}^{|\mathcal{E}|}(\zeta_k-\zeta_k^0)\cdot\mu_k^0.
	\end{equation}


	
	Consider another tension vector $\bar{\boldsymbol{\zeta}}:=(
	\zeta_1,\ldots,\zeta_{|\mathcal{E}|},\zeta^0_{\bar{k}})^T. $
	In order for $\bar{\boldsymbol{\zeta}}$ to be valid, there should exist some $\boldsymbol{y}\in\mathbb{R}^{|\mathcal{V}|}$ such that $\bar{\boldsymbol{\zeta}}=\bar{E}^T\boldsymbol{y}$. 
	%
Since
	\begin{equation}
	\label{eq_tellegen}
	\sum_{k=1}^{|\mathcal{E}|}(\zeta_k-\zeta_k^0)\cdot\mu_k^0=(\bar{\boldsymbol{\zeta}}-\bar{\boldsymbol{\zeta}}^0)^T{\bar{\boldsymbol{\mu}}^0}=(\boldsymbol{y}-\boldsymbol{y}^0)^T{\bar{E}\bar{\boldsymbol{\mu}}^0}=0.
	\end{equation}

	
	Combine~\eqref{eq_cocontentinequality} with~\eqref{eq_tellegen}, and we conclude
	\begin{equation}
	  \label{eq_variationalminimum} \sum_{k=1}^{|\mathcal{E}|}G_k\lvert_{\zeta_k}-\sum_{k=1}^{|\mathcal{E}|}G_k\lvert_{\zeta_k^0}\geq0. 
	\end{equation}
	%
	%
	Inequality~\eqref{eq_variationalminimum} shows that the cocontent of the network system $(\mathcal{G},\Sigma,\Pi)$ reaches its minimum at $(\bar{\boldsymbol{\zeta}}^0,\bar{\boldsymbol{\mu}}^0,\boldsymbol{y}^0)$.
\end{IEEEproof}

\begin{remark}
    The result in~\eqref{eq_tellegen} is known as \emph{Tellegen's Theorem} in circuit theory~\cite{Parodi2018}, which states that if there exist $\boldsymbol{\zeta},\boldsymbol{\mu},\boldsymbol{y}$ such that $E^T\boldsymbol{y}=\boldsymbol{\zeta}$ and $E\boldsymbol{\mu}=\mathbf{0}$, then $\boldsymbol{\mu}^T\boldsymbol{\zeta}=0$.
    Lemma~\ref{lm_cocontent} is a special case of the variational result on the minimum of the cocontent function; see \cite{Parodi2018} for more general results.
\end{remark}

%

%
\begin{example}\label{ex.cocontent}
Consider a simple graph $\mathcal{G}=({\mathcal{V},\mathcal{E}})$ shown in Fig.~\ref{fig_example_minimumheat}, where $\mathcal{V}=\{v_1,v_2,v_3\}$, $\mathcal{E}=\{e_1,e_2\}$.
We add a virtual edge $e_3$ connecting nodes $v_1$ and $v_3$.
The three edges are oriented as the follows: from $v_1$ to $v_2$, from $v_2$ to $v_3$, and from $v_1$ to $v_3$.
The edge functions for $e_1$ and $e_2$ are $\mu_1=\frac{1}{2}\zeta_1$, $\mu_2=\zeta_2$.
\vspace{-10pt}
\begin{figure} [!h]
	\centering
	\includegraphics[height=0.5in]{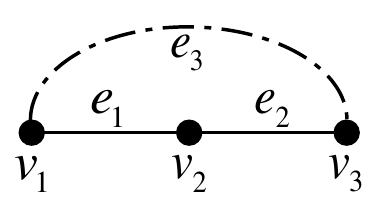}
	\caption{The graph for Example \ref{ex.cocontent}.}
	\label{fig_example_minimumheat}
\end{figure}

	%
	%
	Suppose the value of potential difference between $v_1$ and $v_3$ is $\zeta_3=3$. 
	Lemma~\ref{lm_cocontent} states that the cocontent of the network arrives at its minimum when $\bar{E}\bar{\boldsymbol{\mu}}=\mathbf{0}$, thus, $\mu_1=\mu_2$.
	%
	%
	To make $\bar{E}^T\boldsymbol{y}=\bar{\boldsymbol{\zeta}}$, $\boldsymbol{\mu}=\Psi(\boldsymbol{\zeta})$ and $\bar{E}\bar{\boldsymbol{\mu}}=\mathbf{0}$, we get $\zeta_1=2$, $\zeta_2=1$,  $\mu_1=\mu_2=1$, 
	and the cocontent of the network is $\frac{1}{2}\zeta_1\mu_1+\frac{1}{2}\zeta_2\mu_2=1.5$.
	However, if we let $\zeta_1=1$, $\zeta_2=2$, $\mu_1=0.5$, $\mu_2=2$, such that we can still find valid $\bar{\boldsymbol{\zeta}},\bar{\boldsymbol{\mu}},\boldsymbol{y}$, satisfying $\bar{E}^T\boldsymbol{y}=\bar{\boldsymbol{\zeta}}$, $\boldsymbol{\mu}=\Psi(\boldsymbol{\zeta})$, and $\zeta_3=3$, but $\mu_1\neq\mu_2$, thus $\bar{E}\bar{\boldsymbol{\mu}}\neq\mathbf{0}$. In this case, the cocontent of the network is 2.25, greater than the case when $\mu_1=\mu_2$.
	
	%
\end{example}

Recall the definition of equivalent edge functions proposed in Section~\ref{sec:equivalent}, Lemma~\ref{lm_cocontent} implies that in a strictly positive network if the equivalent edge function exists and the edge functions are all monotonically increasing, the minimum value of the cocontent of the network is exactly the cocontent of the equivalent edge function.
This result is summarized as the following proposition.

\begin{proposition}
    \label{pro_cocontentequivalent}
    Consider a strictly positive network system $(\mathcal{G},\Sigma,\Pi)$ with connected graph  $\mathcal{G}=(\mathcal{V},\mathcal{E}_>)$ represented by~\eqref{eq_plant}-\eqref{eq_umu}, and all edge functions are monotonically increasing. 
	Identify two nodes $p,q\in\mathcal{V}$ as the terminals of interest, and the tension between nodes $p$ and $q$ is specified as $\zeta_{pq}$.
    If the equivalent edge function between nodes $p$ and $q$ exists, then the minimum cocontent of the network system $(\mathcal{G},\Sigma,\Pi)$ is the cocontent of the equivalent edge function between nodes $p$ and $q$, denoted as $\min\mathbf{G}\lvert_{\zeta_{pq}}=G_{pq}\lvert_{\zeta_{pq}}$.
\end{proposition}
\subsection{Convergence Analysis}
With the above formulation, we are now prepared to analyze the signed nonlinear network of nonlinear integrators represented by~\eqref{eq_specialnode},
 \begin{equation}
	\Sigma_i:~~\dot{x}_i(t)=\gamma_i(u_i(t)),~~
	y_i(t)=x_i(t),~~i\in \mathcal{V}.
	\label{eq_specialnode}
\end{equation}
where function $\gamma_i(\cdot)$ satisfies $u_i\cdot\gamma_i(u_i)\geq0$, and the equality holds if and only if $u_i=0$.
It is easy to verify that Assumption~\ref{asp_node} holds for nodes with dynamics~\eqref{eq_specialnode}, and the equilibrium I/O pairs are $\sigma_i=\{(0,\mathbb{R})\}$.
Specifically, if $\gamma_i(u_i)=u_i$, then $\Sigma_i$ is a single integrator.

Now suppose initially we have a strictly positive network of nodes represented by~\eqref{eq_specialnode}. According to Corollary~\ref{thm_strictly}, the nodes' outputs will finally reach agreement.
Now consider a scenario where an attacker is able to add a new negative edge between any two nodes in the original network, or make any existing edge negative.
This can be seen as the attacker adding some disbelief among the group members.
In circuits, the negative edge corresponds to an \emph{ideal Chua's diode}~\cite{Chua1994}, whose current-voltage characteristic is globally active\footnote{A physically realizable Chua's diode is only \emph{locally} active. However, the ideal Chua's diode model is widely used in numerical simulations to investigate the chaotic dynamics~\cite{Recai2010}.}.
We have the following theorem for the convergence property of the network.

\begin{theorem}
	\label{thm_final}
	Consider a signed network system $(\mathcal{G},\Sigma,\Pi)$ with connected graph $\mathcal{G}=(\mathcal{V},\mathcal{E})$ represented by~\eqref{eq_zetay}-\eqref{eq_umu} and~\eqref{eq_specialnode}.	
	Suppose there is only one non-strictly positive edge $\hat{k}$ in $\mathcal{E}$, (i.e., $\forall~k\neq\hat{k}\Rightarrow k\in\mathcal{E}_>$), with its edge function denoted by $\mu_{\hat{k}}(t)=\psi_{\hat{k}}(\zeta_{\hat{k}}(t))$, satisfying $\psi_{\hat{k}}(0)=0$. 
	Furthermore, assume $\forall~k\in\mathcal{E}_>$, $\psi_k(\cdot)$ is monotonically increasing.
	Identify nodes $p$ and $q$, which are connected by edge $\hat{k}$, as the two terminals of the strictly positive subnetwork system $(\mathcal{G}_>,\Sigma,\bar{\Pi})$ with subgraph $\mathcal{G}_>=(\mathcal{V},\mathcal{E}_>)$, and $\bar{\Pi}=\Pi\setminus\{\Pi_{\hat{k}}\}$. 
	If the equivalent edge function between nodes $p$ and $q$ in $(\mathcal{G}_>,\Sigma,\bar{\Pi})$ exists, which we denote as $\bar{\mu}_{pq}(t)=\bar{\psi}_{pq}(\zeta_{\hat{k}}(t))$,
	and
	%
	$$(\mu_{\hat{k}}(t)+\bar{\mu}_{pq}(t))\cdot\zeta_{\hat{k}}(t)\geq0$$ holds for any $\zeta_{\hat{k}}(t)\in\mathbb{R}$, then $\lim\limits_{t\rightarrow\infty}\boldsymbol{u}(t)=\mathbf{0}$,
	and $\lim\limits_{t\to\infty}\mu_{\hat{k}}(t)+\bar{\mu}_{pq}(t)=0$.
\end{theorem}
\begin{IEEEproof}
	Let $V(t)$ be the cocontent of the network system $(\mathcal{G},\Sigma,\Pi)$. Then $V(t)=G_{\hat{k}}\lvert_{\zeta_{\hat{k}}(t)}+\mathbf{G}_>\lvert_{\zeta_{\hat{k}}(t)}$, where $G_{\hat{k}}\lvert_{\zeta_{\hat{k}}(t)}$ is the cocontent of edge $\hat{k}$, and $\mathbf{G}_>\lvert_{\zeta_{\hat{k}}(t)}$ is the cocontent of the subnetwork system $(\mathcal{G}_>,\Sigma,\bar{\Pi})$ for a fixed value of $\zeta_{\hat{k}}(t)$.
	
	We first show that $V(t)\geq0$.
	Since $\forall~k\neq\hat{k},k\in\mathcal{E}$, $\psi_k(\cdot)$ is monotonically increasing, according to Proposition~\ref{pro_cocontentequivalent}, 
	$\mathbf{G}_>\lvert_{\zeta_{\hat{k}}(t)}\geq G_{pq}\lvert_{\zeta_{\hat{k}}(t)}$, where $G_{pq}\lvert_{\zeta_{\hat{k}}(t)}$ is the concontent of the equivalent edge function of the strictly positive subnetwork system $(\mathcal{G}_>,\Sigma,\bar{\Pi})$ when $\zeta_{\hat{k}}(t)$ is specified.
	%
	%
	If $(\mu_{\hat{k}}(t)+\bar{\mu}_{pq}(t))\cdot\zeta_{\hat{k}}(t)\geq0$ holds for any $\zeta_{\hat{k}}(t)\in\mathbb{R}$, recalling the relationship between the edge function and the cocontent function shown in Fig.~\ref{fig_cocontent}, we obtain $G_{\hat{k}}\lvert_{\zeta_{\hat{k}}(t)}+G_{pq}\lvert_{\zeta_{\hat{k}}(t)}\geq 0$.
	Therefore, $V(t)=G_{\hat{k}}\lvert_{\zeta_{\hat{k}}(t)}+\mathbf{G}_>\lvert_{\zeta_{\hat{k}}(t)}\geq0$.
	
	As $\dot{G}_k\lvert_{\zeta(t)}(t)=\mu_k(t)\dot{\zeta}_k(t),~\forall~k\in\mathcal{E}$, then
	$$
	\begin{aligned}
	\dot{V}(t)&=\boldsymbol{\mu}(t)^T\dot{\boldsymbol{\zeta}}(t)=\boldsymbol{\mu}(t)^TE^T\dot{\boldsymbol{y}}(t)=-\boldsymbol{u}(t)^T\dot{\boldsymbol{y}}(t)\\
	&=-\sum_{i=1}^{|\mathcal{V}|}u_i(t)\cdot\gamma_i(u_i(t))\leq0.
	\end{aligned}
	$$
	From LaSalle's invariance principle~\cite{Hassan}, the system will converge to the largest invariant set satisfying  $\lim\limits_{t\rightarrow\infty}\boldsymbol{u}(t)=\mathbf{0}$.
	In that case, in order to satisfy~\eqref{eq_umu}, the flow into the two-terminal subnetwork system $(\mathcal{G}_>,\Sigma,\bar{\Pi})$ should equal the negative of the flow on edge $\hat{k}$, i.e., $\lim\limits_{t\to\infty}\mu_{\hat{k}}(t)+\bar{\mu}_{pq}(t)=0$ holds.
\end{IEEEproof}

Theorem~\ref{thm_final} shows that the signed nonlinear network forms one or several steady clusters, if the sum of the equivalent edge function of the strictly positive network and the non-positive edge function is passive. 
%
Under such a condition, $\zeta_{\hat{k}}$ will converge to a point satisfying $\lim\limits_{t\to\infty}\mu_{\hat{k}}(t)+\bar{\mu}_{pq}(t)=0$.
Since all the other edges except $\hat{k}$ are strictly positive,
it follows from Proposition~\ref{pro_clustermax} that the nodes incident to only strictly positive edges cannot be the only node with the maximum or minimum steady output. Thus we conclude $\lim\limits_{t\to\infty}|\zeta_k(t)|\leq \lim\limits_{t\to\infty}|\zeta_{\hat{k}}(t)|$, $\forall~k\neq\hat{k}$.
%
The following corollary is a direct result of Theorem~\ref{thm_final}, which provides a sufficient condition for the network reaching agreement.
\begin{corollary}
	\label{corollary_final}
	Consider a signed network system $(\mathcal{G},\Sigma,\Pi)$ with connected graph $\mathcal{G}=(\mathcal{V},\mathcal{E})$ represented by~\eqref{eq_zetay}-\eqref{eq_umu} and~\eqref{eq_specialnode}.	
	Suppose there is only one non-strictly positive edge $\hat{k}$ in $\mathcal{E}$, (i.e., $\forall~k\neq\hat{k}\Rightarrow k\in\mathcal{E}_>$), with its edge function denoted by $\mu_{\hat{k}}(t)=\psi_{\hat{k}}(\zeta_{\hat{k}}(t))$, satisfying $\psi_{\hat{k}}(0)=0$. 
	Furthermore, assume $\forall~k\in\mathcal{E}_>$, $\psi_k(\cdot)$ is monotonically increasing.
	Identify nodes $p$ and $q$, which are connected by edge $\hat{k}$, as the two terminals of the strictly positive subnetwork system $(\mathcal{G}_>,\Sigma,\bar{\Pi})$ with subgraph $\mathcal{G}_>=(\mathcal{V},\mathcal{E}_>)$, and $\bar{\Pi}=\Pi\setminus\{\Pi_{\hat{k}}\}$. 
	If the equivalent edge function between nodes $p$ and $q$ in $(\mathcal{G}_>,\Sigma,\bar{\Pi})$ exists, which we denote as $\bar{\mu}_{pq}(t)=\bar{\psi}_{pq}(\zeta_{\hat{k}}(t))$,
	and
	%
	$$(\mu_{\hat{k}}(t)+\bar{\mu}_{pq}(t))\cdot\zeta_{\hat{k}}(t)\geq0$$ holds for any $\zeta_{\hat{k}}(t)\in\mathbb{R}$, 
 and $(\mu_{\hat{k}}(t)+\bar{\mu}_{pq}(t))\cdot\zeta_{\hat{k}}(t) = 0$ only if $\zeta_{\hat{k}}(t)=0$, then $\lim\limits_{t\rightarrow\infty}\boldsymbol{\zeta}(t)=\mathbf{0}$, and $\lim\limits_{t\rightarrow\infty}\boldsymbol{y}(t)=\beta\mathbf{1},~\beta\in\mathbb{R}$.
	%
\end{corollary}
\begin{remark}
	Theorem~\ref{thm_final} is a generalization of Theorem III.3 in~\cite{Zelazo2014} for signed linear networks of single integrators with scalar weights to the nonlinear case.
	If all edge functions are in the form of~\eqref{eq_linear},
	the network can reach agreement as long as the scalar weight $w_{\hat{k}}$ for the only one negative edge $\hat{k}$ satisfies $|w_{\hat{k}}|<\frac{1}{\bar{r}_{pq}}$, where $p,q$ are the nodes connected by $\hat{k}$, and $\bar{r}_{pq}$ is the effective resistance of the two-terminal network without edge $\hat{k}$.
	In this scenario, the equivalent edge function of the two-terminal network is $\bar{\mu}_{pq}(t)=\zeta_{\hat{k}}(t)/\bar{r}_{pq}$, and $(\mu_{\hat{k}}(t)+\bar{\mu}_{pq}(t))\cdot\zeta_{\hat{k}}(t)\geq0$ holds for any $\zeta_{\hat{k}}(t)\in\mathbb{R}$, with the equality holds only if $\zeta_{\hat{k}}(t)=0$.
	Therefore, $\lim\limits_{t\rightarrow\infty}\boldsymbol{\zeta}(t)=\mathbf{0}$, meaning the all nodes finally reach agreement.	
	On the other hand, if the negative edge weight $w_{\hat{k}}=-\frac{1}{\bar{r}_{pq}}$, then $(\mu_{\hat{k}}(t)+\bar{\mu}_{pq}(t))\cdot\zeta_{\hat{k}}(t)=0$ holds for any $\zeta_{\hat{k}}(t)\in\mathbb{R}$.
	In this case, we can still guarantee the convergence of the network, however, $\boldsymbol{\zeta}(t)$ does not necessarily converge to $\mathbf{0}$, meaning we can get clustering result.
\end{remark}

	As a generalization of Theorem~III.4 in~\cite{Zelazo2014}, we are able to extend the convergence criteria to the nonlinear case where any two non-strictly positive edges are not contained in the same cycle. Note that according to the definition of equivalent edge function, it is required to satisfy $\bar{E}\bar{\boldsymbol{\mu}}=\mathbf{0}$. Since the null space of $\bar{E}$ is spanned by all the linearly independent \emph{signed path vectors} 
 corresponding to the cycles~\cite{Godsil}, only the edges that are contained in the same cycle with $\hat{k}$ can influence the equivalent edge function $\bar{\psi}_{pq}$, where $p$ and $q$ are the nodes connected by $\hat{k}$.
	Thus we have the following corollary.

\begin{corollary}	
	Consider a signed network system $(\mathcal{G},\Sigma,\Pi)$ with connected graph $\mathcal{G}=(\mathcal{V},\mathcal{E})$ represented by~\eqref{eq_zetay}-\eqref{eq_umu} and~\eqref{eq_specialnode}.	
	Denote the set of non-strictly positive edges as $\hat{\mathcal{E}}$ (i.e., $\mathcal{E}=\hat{\mathcal{E}}\cup \mathcal{E}_>$), and suppose any two edges in $\hat{\mathcal{E}}$ are not contained in the same cycle.
	For any $\hat{k}\in \hat{\mathcal{E}}$, its edge function is denoted by $\mu_{\hat{k}}(t)=\psi_{\hat{k}}(\zeta_{\hat{k}}(t))$,
	satisfying $\psi_{\hat{k}}(0)=0$. 
	Furthermore, assume $\forall~k\in\mathcal{E}_>$, $\psi_k(\cdot)$ is monotonically increasing.
	For all $\hat{k}\in \hat{\mathcal{E}}$,	
	identify nodes $p_{\hat{k}}$ and $q_{\hat{k}}$, which are connected by edge $\hat{k}$, as a pair of the two terminals for the strictly positive subnetwork system $(\mathcal{G}_>,\Sigma,\Pi\setminus\{\Pi_{\hat{k}}\colon \hat{k}\in \hat{\mathcal{E}}\})$ with subgraph $\mathcal{G}_>=(\mathcal{V},\mathcal{E}_>)$. 
	If the following two conditions hold for all $\hat{k}\in \hat{\mathcal{E}}$,
	\begin{itemize}
		\item[i)]  the equivalent edge function between nodes $p_{\hat{k}}$ and $q_{\hat{k}}$ in $(\mathcal{G}_>,\Sigma,\Pi\setminus\{\Pi_{\hat{k}}\colon \hat{k}\in \hat{\mathcal{E}}\})$ exists, which we denote as $\bar{\mu}_{p_{\hat{k}}q_{\hat{k}}}(t)=\bar{\psi}_{p_{\hat{k}}q_{\hat{k}}}(\zeta_{\hat{k}}(t))$;
		\item[ii)] $(\mu_{\hat{k}}(t)+\bar{\mu}_{p_{\hat{k}}q_{\hat{k}}}(t))\cdot\zeta_{\hat{k}}(t)\geq0$ holds for any $\zeta_{\hat{k}}(t)\in\mathbb{R}$;
	\end{itemize}
	%
	then $\lim\limits_{t\rightarrow\infty}\boldsymbol{u}(t)=\mathbf{0}$,
	and $\lim\limits_{t\to\infty}\mu_{\hat{k}}(t)+\bar{\mu}_{p_{\hat{k}}q_{\hat{k}}}(t)=0,~\forall~\hat{k}\in \hat{\mathcal{E}}$.
\end{corollary}

\subsection{Clustering Analysis}
%
%
%

%

Now we generalize the specific clustering scenario of signed linear networks discussed in Proposition IV.1 of~\cite{Zelazo2014}, where there is only one single cycle and one single non-strictly positive edge in the network, and all edge functions are in the form of~\eqref{eq_linear}.  The result showed
that, if the edge weight of the non-strictly positive edge equals the negative inverse of the effective resistance of the remaining two-terminal network, then the number of the clusters equals the number of nodes on the cycle.
%
%
We also show that the result even holds with more cycles in the network, as long as there is only one cycle containing the non-strictly positive edge.
We first provide the following proposition.
\begin{proposition}
	\label{pro_clusternumber}
	Consider a signed network system $(\mathcal{G},\Sigma,\Pi)$ with connected graph $\mathcal{G}=(\mathcal{V},\mathcal{E})$ represented by~\eqref{eq_plant}-\eqref{eq_umu} and suppose Assumption~\ref{asp_node} holds.
	Suppose there is only one non-strictly positive edge $\hat{k}$ in $\mathcal{E}$ (i.e., $\forall~k\neq\hat{k}\Rightarrow k\in\mathcal{E}_>$). 
	Assume there is only one cycle in $\mathcal{G}$ containing edge $\hat{k}$.
	If $\lim\limits_{t\to\infty}\boldsymbol{u}(t)=\mathbf{0}$, and $\lim\limits_{t\to\infty}\boldsymbol{\zeta}(t)\neq\mathbf{0}$, 
	then the number of clusters formed by the nodes' outputs equals the number of nodes in the cycle containing edge $\hat{k}$.
\end{proposition}
\begin{IEEEproof}
	%
	Since only one cycle (denoted as $\mathcal{C}$) contains edge $\hat{k}$,
	if we remove all the edges in $\mathcal{C}$, then the number of connected components equals the number of nodes in $\mathcal{C}$, and each component contains exactly one node of $\mathcal{C}$.
	%
	%
	In each component, since there is at most one node not only incident to strictly positive edges, from Proposition~\ref{pro_clustermax}, we conclude that nodes' outputs in each component will reach agreement.
	Now we show that if there are more than one clusters formed, then the nodes in $\mathcal{C}$ have different output values. 
	Denote the two nodes connected by $\hat{k}$ are nodes $p$ and $q$.
	According to Corollary~\ref{coro_signedstrictly}, if $\lim\limits_{t\to\infty}y_p(t)-y_q(t)=0$, then $\lim\limits_{t\to\infty}\boldsymbol{\zeta}(t)=\mathbf{0}$. 
	Now suppose $\lim\limits_{t\to\infty}y_p(t)-y_q(t)>0$.
	Since there is only one non-strictly positive edge in $\mathcal{G}$, it can be concluded from Proposition~\ref{pro_clustermax} that $\lim\limits_{t\to\infty}y_p(t)\geq\lim\limits_{t\to\infty}y_i(t)\geq\lim\limits_{t\to\infty}y_q(t),~\forall~i\neq p,q,~i\in\mathcal{V}$.
	Consider node $p'$ in cycle $\mathcal{C}$, who is a neighbor of node $p$, then $\lim\limits_{t\to\infty}y_{p'}(t)-y_p(t)\leq0$. 
	However, if $\lim\limits_{t\to\infty}y_{p'}(t)-y_p(t)=0$, with Proposition~\ref{pro_clustermax}, proceeding forward, we get $\lim\limits_{t\to\infty}y_p(t)=\lim\limits_{t\to\infty}y_{p'}(t)=\ldots=\lim\limits_{t\to\infty}y_q(t)$, which contradicts $\lim\limits_{t\to\infty}y_p(t)-y_q(t)>0$.
	Therefore, $\lim\limits_{t\to\infty}y_{p'}(t)-y_p(t)<0$, proceeding forward, we get $\lim\limits_{t\to\infty}y_p(t)>\lim\limits_{t\to\infty}y_{p'}(t)>\ldots>\lim\limits_{t\to\infty}y_q(t)$, meaning the outputs of the nodes in $\mathcal{C}$ differ from each other.
	
	Therefore, we can conclude that the number of clusters equals the length of cycle $\mathcal{C}$, and the outputs of the nodes in $\mathcal{C}$ form a decreasing or an increasing sequence along the path consisting of only strictly positive edges in $\mathcal{C}$.
\end{IEEEproof}

By combining Theorem~\ref{thm_final} and Proposition~\ref{pro_clusternumber}, we can generalize Proposition IV.1 in~\cite{Zelazo2014} as the following corollary. 
\begin{corollary}
	\label{coro_cyclefinal}
	Consider a signed network system $(\mathcal{G},\Sigma,\Pi)$ with connected graph $\mathcal{G}=(\mathcal{V},\mathcal{E})$ represented by~\eqref{eq_zetay}-\eqref{eq_umu} and~\eqref{eq_specialnode}.	
	Suppose there is only one non-strictly positive edge $\hat{k}$ in $\mathcal{E}$, (i.e., $\forall~k\neq\hat{k}\Rightarrow k\in\mathcal{E}_>$), with its edge function denoted by $\mu_{\hat{k}}(t)=\psi_{\hat{k}}(\zeta_{\hat{k}}(t))$, satisfying $\psi_{\hat{k}}(0)=0$. 
	Furthermore, assume $\forall~k\in\mathcal{E}_>$, $\psi_k(\cdot)$ is monotonically increasing.
	 Identify nodes $p$ and $q$, which are connected by edge $\hat{k}$, as the two terminals of the strictly positive subnetwork system $(\mathcal{G}_>,\Sigma,\bar{\Pi})$ with subgraph $\mathcal{G}_>=(\mathcal{V},\mathcal{E}_>)$, and $\bar{\Pi}=\Pi\setminus\{\Pi_{\hat{k}}\}$. 
	If the equivalent edge function between nodes $p$ and $q$ in $(\mathcal{G}_>,\Sigma,\bar{\Pi})$ exists, which we denote as $\bar{\mu}_{pq}(t)=\bar{\psi}_{pq}(\zeta_{\hat{k}}(t))$,
	and
	%
	$(\mu_{\hat{k}}(t)+\bar{\mu}_{pq}(t))\cdot\zeta_{\hat{k}}(t)\geq0$ holds for any $\zeta_{\hat{k}}(t)\in\mathbb{R}$, and there is only one cycle in $\mathcal{G}$ containing edge $\hat{k}$, then
	\begin{itemize}
		\item[i)] $\lim\limits_{t\rightarrow\infty}\boldsymbol{u}(t)=\mathbf{0}$;
		\item[ii)] $\lim\limits_{t\to\infty}\mu_{\hat{k}}(t)+\bar{\mu}_{pq}(t)=0$;
		\item[iii)] the number of clusters formed by the nodes' outputs is either one, or length of the  cycle containing edge $\hat{k}$.
	\end{itemize}
\end{corollary}
\section{Simulation Results\label{sec:simulation}}	

\begin{figure} [!t]
	\centering
	\includegraphics[height=1.2in]{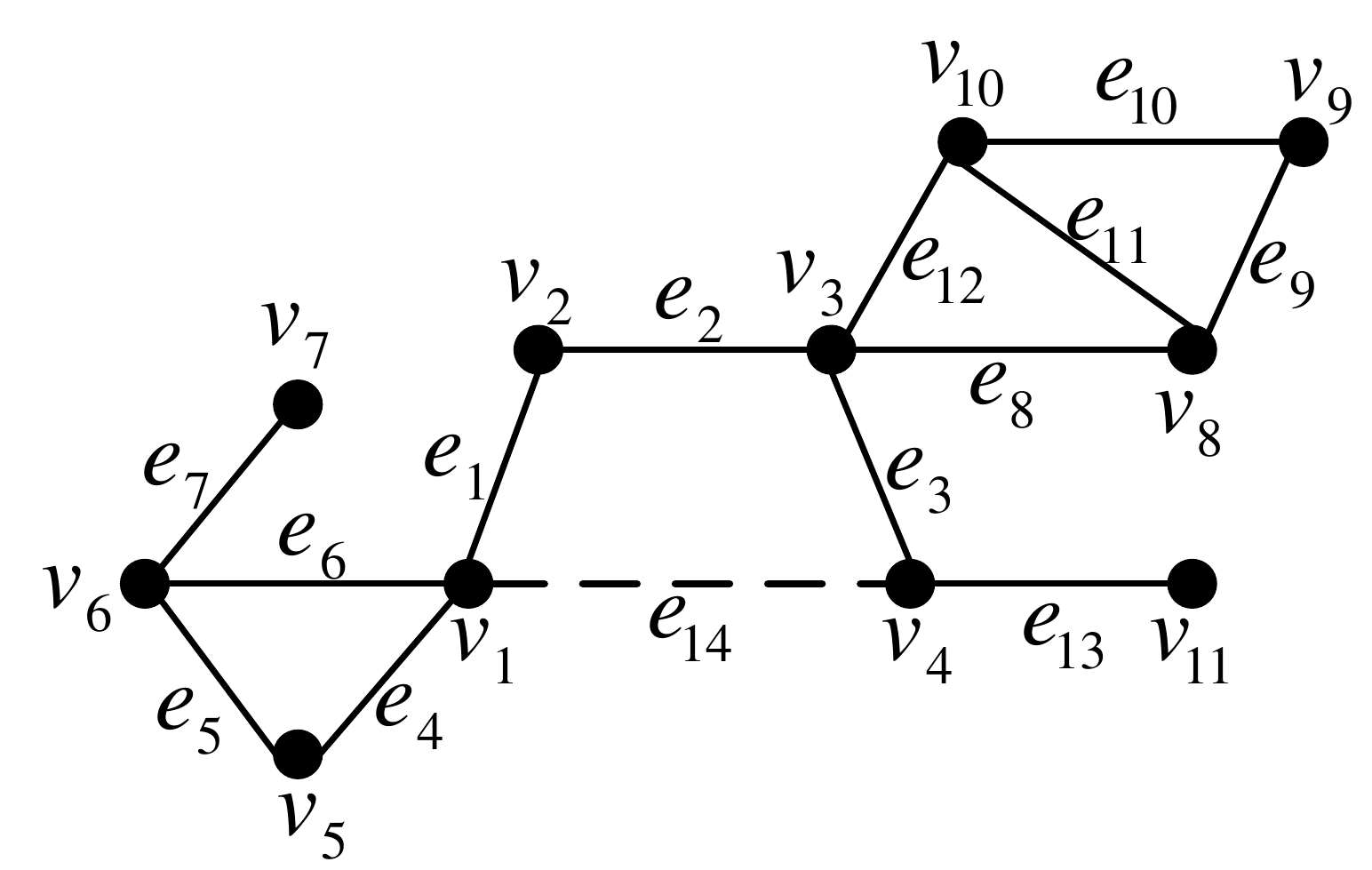}
	\vspace{-10pt}
	\caption{A network of 11 nodes and 14 edges. Edge $e_{14}$ is introduced by an attacker. The other 13 edges are strictly positive, while $e_{14}$ is strictly negative.}
	\label{fig_manycycle}
\end{figure}

We present a numerical simulation to illustrate the main results of this paper.
Consider a network of single integrators shown in Fig.~\ref{fig_manycycle}. 
The original network consists of 11 nodes and 13 edges (the edges of the original network are represented by solid lines in Fig.~\ref{fig_manycycle}).
All the 13 edges are strictly positive, and their edge functions are described by~\eqref{eq_finitetime}, which is used in~\cite{WangLong2010} to achieve finite-time consensus of integrators,
\begin{equation}
\label{eq_finitetime}
	\mu_k(t)=w_k\cdot\mathrm{sign}(\zeta_k(t))\cdot|\zeta_k(t)|^{\alpha_k},
\end{equation}
where $w_k>0$, $0<\alpha_k<1$. 
We use $\boldsymbol{w}$ and $\boldsymbol{\alpha}$ to represent the stacked vector of $w_k$ and $\alpha_k$, respectively, and we set the parameters as $\boldsymbol{w}=(3,2,4,1,2,1,3,2,2,1,1,1,2)^T$, $\boldsymbol{\alpha}=(0.4,0.5,0.2,0.8,0.4,0.4,0.5,0.5,0.5,0.6,0.8,0.2,0.5)^T$.

Now we add a strictly negative edge between nodes $v_1$ and $v_4$, which is labeled as $e_{14}$, and shown by dashed line in Fig.~\ref{fig_manycycle}. One can easily translate the network shown in Fig.~\ref{fig_manycycle} into its corresponding circuit model similar to Fig.~\ref{fig_circuit_circuit}, where each node corresponds to a capacitor connected to the ground with the capacitance being 1 Farad, and each edge corresponds to a resistor, i.e., the edges in the form of~\eqref{eq_finitetime} represent general passive resistors, while the negative edge $e_{14}$ represents an ideal Chua's diode.
%
Only one cycle contains $e_{14}$ in the network, that is, the cycle consisting of nodes $v_1$, $v_2$, $v_3$ and $v_4$, and edges $e_1$, $e_2$, $e_3$ and $e_{14}$.

	\begin{figure*}[!htb]
		\centering
		\subfigure[]
		{\includegraphics[height=1.2in]{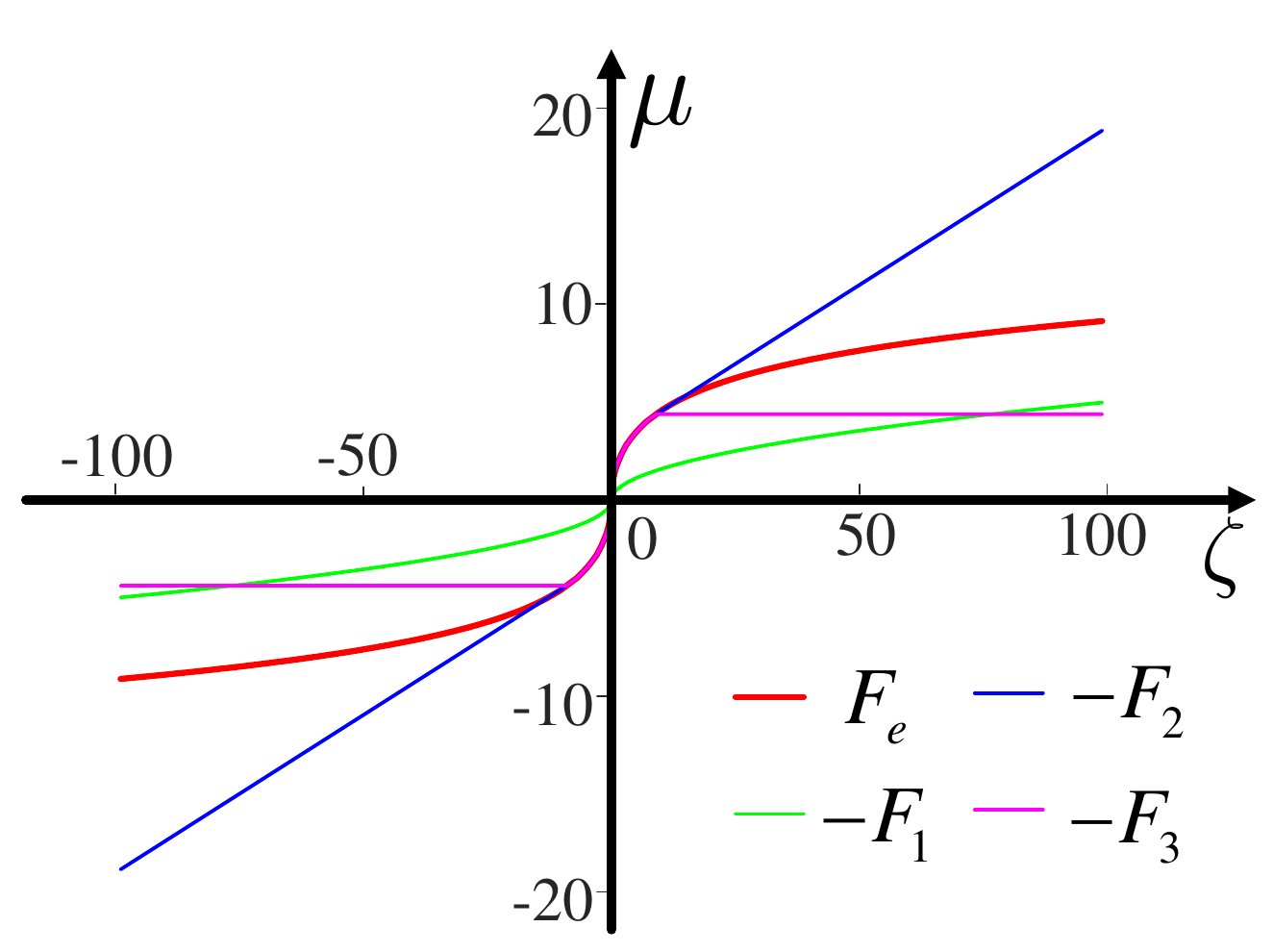}
			\label{fig_sub1a}}
		\subfigure[]{
			\includegraphics[height=1.2in]{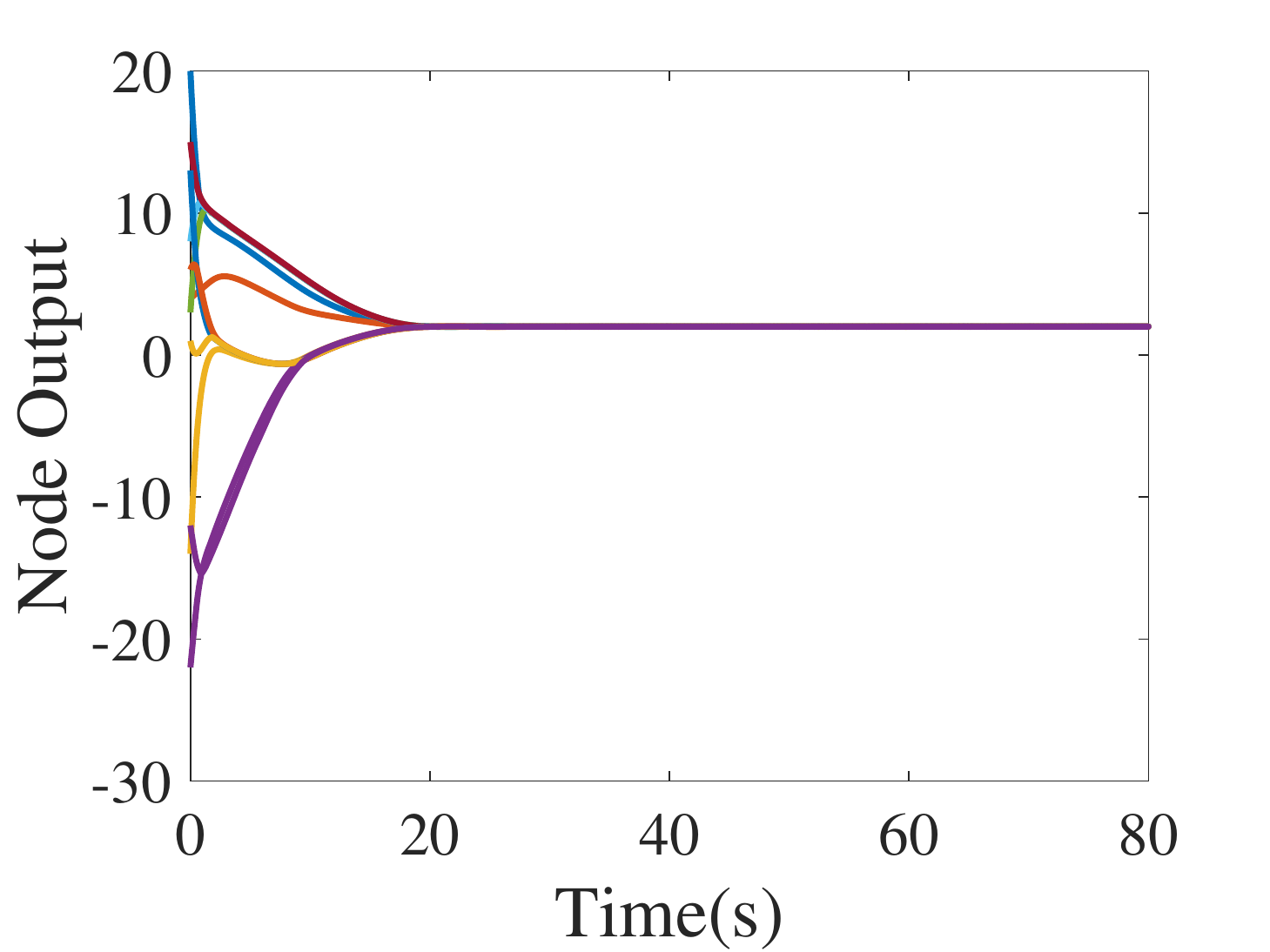}
			\label{fig_sub1b}}	
		\subfigure[]{
			\includegraphics[height=1.2in]{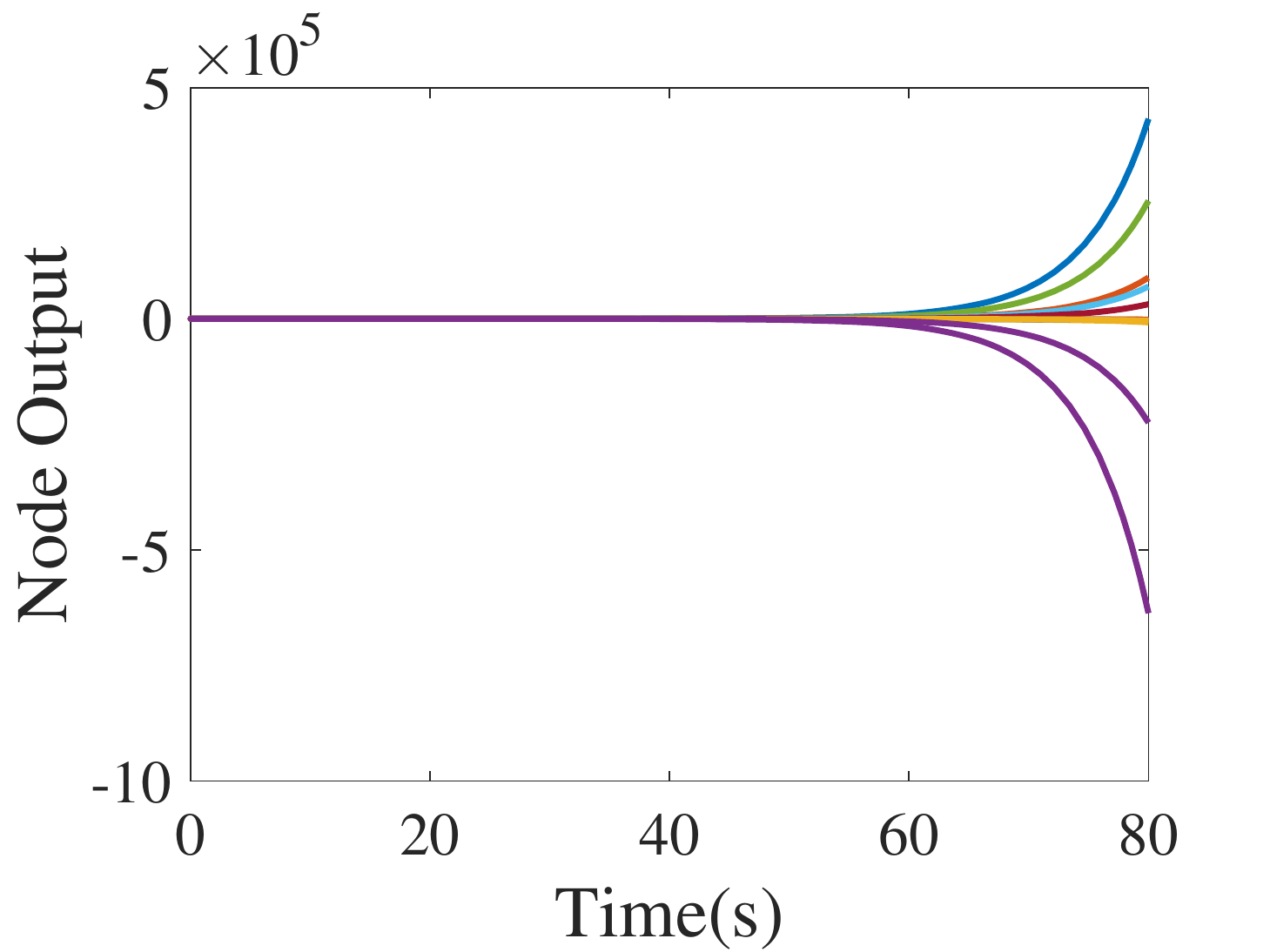}
			\label{fig_sub1c}}	
		\subfigure[]{
			\includegraphics[height=1.2in]{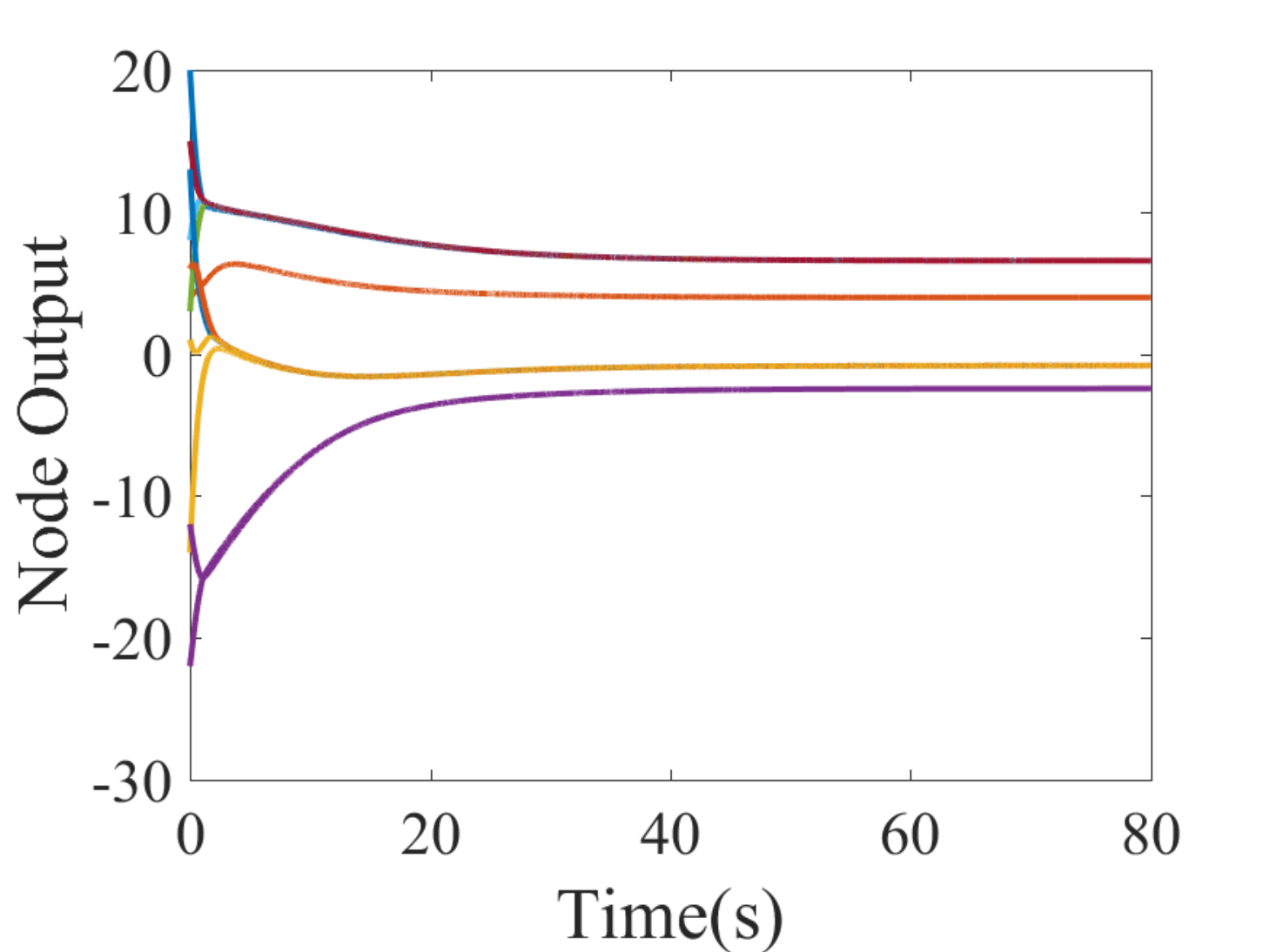}
			\label{fig_sub1d}}	
			\vspace{-5pt}				
		\caption{Simulation results of the network in Fig.~\ref{fig_manycycle}, when different edge functions are chosen for $e_{14}$. (a) Equivalent edge function ($F_e$) of the original two-terminal network between $v_1$ and $v_4$, and the opposite of the three candidate edge functions for $e_{14}$ (the three candidate edge functions are denoted as $F_1$, $F_2$, and $F_3$.). (b) Agreement result when the edge function for $e_{14}$ is chosen as $F_1$. (c) Divergence result when the edge function for $e_{14}$ is chosen as $F_2$. (d) Clustering result when the edge function for $e_{14}$ is chosen as $F_3$.}
		\label{fig_simuresult}

	\end{figure*}
	
%

%
%

We use Algorithm~\ref{alg} to approximate the equivalent edge function of the original two-terminal network between nodes $v_1$ and $v_4$.
We obtain $\{\zeta_{14}\}$ by sampling in $[-100,100]$, 
and the algorithm proposed in~\cite{Noda2011} is used to calculate the operating point of the corresponding circuit (Line~\ref{ln_calculate} in Algorithm~\ref{alg}).

%
%
Since all the edge functions of the original network are monotonically increasing, and $\mu_k(t)\to\pm\infty$ as $\zeta_k(t)\to\pm\infty$, according to Proposition~\ref{pro_existence}, the equivalent edge function of the two terminal network exists.
The caluclated equivalent edge function, $F_e$ of the two-terminal network between nodes $v_1$ and $v_4$ is shown in Fig.~\ref{fig_sub1a}.

We now consider three strictly negative candidate functions for edge $e_{14}$, i.e., $F_1$, $F_2$ and $F_3$, and we show their opposites, i.e., $-F_1$, $-F_2$ and $-F_3$ in Fig.~\ref{fig_sub1a}.
The initial states under three different conditions are set as the same, which are $\boldsymbol{x}(0)=[20,4,-14,-22,3,8,15,13,6,1,-12]^T$.
We execute the interaction protocols respectively, and show the simulation results in Fig.~\ref{fig_sub1b}-\ref{fig_sub1d}.

In Fig.~\ref{fig_sub1a}, it can be seen that $F_e+F_1$ is still input strictly passive, as a result, when the negative edge function is $F_1$, all nodes will still converge to the agreement space, as shown in Fig.~\ref{fig_sub1b}, demonstrating Corollary~\ref{corollary_final}.
In the second case, $F_e+F_2$ is active, and the outputs of the integrators will diverge, as shown in Fig.~\ref{fig_sub1c}.
In the third case, $F_e+F_3$ is passive, but not input strictly passive, and its equilibria is $[-9,9]$.
We see the outputs of the integrators form clusters in this case.
Besides, in the clustering scenario, there are exactly four clusters.
The steady outputs of cluster with nodes $v_1$, $v_5$, $v_6$, and $v_7$ are the maximum at around 6.55, and the steady outputs of cluster with nodes $v_4$ and $v_{11}$ are the minimum at around -2.45, the distance between these two steady clusters is around 9, which is the boundary of the equilibria of $F_e+F_3$, and thus demonstrating Corollary~\ref{coro_cyclefinal}.

\section{Concluding Remarks\label{sec:conclusion}}
This work explored a nonlinear extension to the notion of signed networks.  For a broad class of network systems, we provided results on the agreement and clustering phenomena that may be observed.  We note that to reach agreement we require a spanning subgraph of strictly positive edges.  For the case of integrator agents, we also proposed a nonlinear interpretation for the effective resistance of a network, and used that to provide sufficient conditions for convergence of these networks with negative edges.  

We believe this work to be an important first step towards a more general theory of nonlinear signed networks.  Open questions that remain include expanding our convergence analysis to include edge functions that are dynamic.  Along these lines, we must also develop equivalent functions for such edges, and for networks comprised of general MEIP nodes.  These are subjects of our future works.  

\balance
\bibliographystyle{IEEEtran}
\bibliography{jfrExampleRefs}

\end{document}